\documentclass[12pt,onecolumn,aps,pra,floatfix,superscriptaddress,preprint, showpacs,groupedaddress]{revtex4-1}

\usepackage{times}
\usepackage{graphicx}
\usepackage{graphics}
\usepackage{amssymb}
\usepackage{amsmath}
\usepackage{epsfig}
\usepackage{latexsym}
\usepackage{color}
\usepackage{rotating}
\usepackage{subfigure}
\usepackage{float}
\usepackage[breaklinks=true]{hyperref}

\usepackage{setspace}

\hypersetup{
  colorlinks   = true, %Colours links instead of ugly boxes
  urlcolor     = blue, %Colour for external hyperlinks
  linkcolor    = blue, %Colour of internal links
  citecolor   =  red %Colour of citations
}

\usepackage{soul} % for strikethrough \st

\begin{document}

\title{Quantum correlations overcome the photodamage limits of light microscopy}

\author{Catxere A. Casacio$^1$, Lars S. Madsen$^1$, Alex Terrasson$^1$, Muhammad Waleed$^1$, Kai Barnscheidt$^2$, Boris Hage$^2$, Michael A. Taylor$^3$, and Warwick P. Bowen$^{1,}$\footnote{w.bowen@uq.edu.au}}
\affiliation{$^1$ARC Centre of Excellence for Engineered Quantum Systems, University of Queensland, St Lucia, QLD 4072, Australia}
\affiliation{$^2$Institut f{\"u}r Physik, Universit{\"a}t Rostock, Rostock, Germany}
\affiliation{$^3$Australian Institute for Bioengineering and Nanotechnology, The University of Queensland, St Lucia, QLD 4072, Australia}

%\date{}

\baselineskip24pt

\begin{abstract}
State-of-the-art microscopes use intense lasers that can severely disturb biological processes, function and viability. This introduces hard  limits on performance that only quantum photon correlations can overcome. Here we demonstrate this absolute quantum advantage, achieving signal-to-noise beyond the photodamage-free capacity of conventional microscopy. We achieve this in a coherent Raman microscope, which we use to image molecular bonds within a cell with both quantum-enhanced contrast and sub-wavelength resolution. This allows the observation of nanoscale biological structures that would otherwise not be resolved. Coherent Raman microscopes allow highly selective biomolecular finger-printing in unlabelled specimens, but photodamage is a major roadblock for many applications. By showing that this roadblock can be overcome, our work  provides a path towards order-of-magnitude improvements in both sensitivity and imaging speed.
\end{abstract}

\maketitle

\section*{Introduction}

Light microscopy has a long tradition of providing deep insights into the nature of living systems. 
Recent advances range from super-resolution microscopes that allow the imaging of biomolecules at near atomistic resolution~\cite{sigal2018visualizing}, to light-sheet techniques that rapidly explore living cells in three-dimensions~\cite{valm2017applying}, and adaptive high-speed microscopes for optogenetic control of neural networks~\cite{nadella2016random,adam2019voltage}. The performance of these microscopes is limited by the stochastic nature of light --- that it exists in discrete packets of energy, i.e. photons. Randomness in the times that photons are detected introduces shot-noise, fundamentally constraining sensitivity, resolution and speed~\cite{taylor2016quantum}. The long-established solution to this problem is to increase the intensity of the illumination light. However, for many advanced microscopes  this approach is no longer tenable due to the intrusion of the light on biological processes~\cite{li2020adaptive,fu2006characterization, schermelleh2019super}.
Light is known to disturb function, structure and growth~\cite{schermelleh2019super, fu2006characterization, waldchen2015light}, and is ultimately fatal~\cite{waldchen2015light, fu2006characterization}. 

It has been known for many decades that quantum correlations can be used to extract more information per photon used in an optical measurement~\cite{Slusher}. This allows the trade-off between  signal-to-noise and damage to be broken~\cite{sewell2013certified}. Indeed, for this reason quantum correlations are now used routinely to improve the performance of laser interferometric gravitational wave detectors~\cite{aasi2013enhanced}. They have also been shown to improve many other optical measurements in proof-of-principle experiments~\cite{giovannetti2011advances}.
The importance of addressing biological photodamage 
has motivated efforts to apply quantum-correlated illumination into microscopy, with recent demonstrations of quantum-enhanced absorption~\cite{brida2010experimental, defienne2019quantum, sabines2019twin, samantaray2017realization} and phase-contrast~\cite{israel2014supersensitive, ono2013entanglement} imaging.  Quantum correlations 
have also been used for illumination in infrared spectroscopic imaging~\cite{kalashnikov2016infrared} and optical coherence tomography~\cite{paterova2018tunable}.  However, all previous experiments used optical intensities more than twelve orders of magnitude lower than those for which biophysical damage typically arises~\cite{mauranyapin2017evanescent},  and therefore did not provide an absolute sensitivity advantage -- superior sensitivity could have been achieved in the absence of quantum correlations using higher optical power.
Increasing the illumination intensity to levels relevant for high performance microscopy is a longstanding challenge, 
 that has proved difficult due to limitations in methods used to produce quantum correlations, to their fragility once produced, and to the challenge of integration within a precision microscope. 

Here we report a microscope that operates safely with signal-to-noise beyond the damage limit of coherent illumination. 
Light-induced damage is directly observed, imposing 
 a hard bound on  intensity, and therefore signal-to-noise.
We overcome this bound using quantum-correlated light
and apply the technique to intracellular imaging with quantum-enhanced contrast and sub-wavelength resolution.
This allows nanoscale biological features to be seen that would have otherwise been buried beneath the shot-noise.
While the concept is applicable broadly in precision microscopy, we implement it here in a coherent Raman scattering microscope \cite{camp2015chemically, cheng2015vibrational, hu2019biological}. Coherent Raman microscopes probe the vibrational spectra of biomolecules, allowing unlabelled imaging of chemical bonds with exceptionally high specificity -- far higher than is possible, for example, using fluorescence \cite{hu2019biological, cheng2015vibrational, camp2014high,wei2017super}.
This provides new capabilities to study a wide range of biosystems and processes, including
neurotransmitters~\cite{fu2017label}, 
 metabolic processes~\cite{zhang2019spectral}, nerve degeneration~\cite{tian2016monitoring}, neuron membrane potentials~\cite{LiuBin} and antibiotic response~\cite{schiessl2019phenazine}. However, photodamage places acute constraints on performance~\cite{camp2015chemically, cheng2015vibrational, fu2006characterization}, presenting a roadblock for powerful prospective applications such as label-free spectrally-multiplexed imaging~\cite{camp2014high,wei2017super}.
State-of-the-art coherent Raman microscopes are already limited by shot-noise \cite{saar2010video,freudiger2014stimulated}. The roadblock therefore cannot be overcome through improvements in instrumentation. By using quantum correlations to overcome it,
 our results remove a fundamental barrier  to advances in coherent Raman microscopy and high performance microscopy more broadly.

\section*{Quantum-compatible coherent Raman microscope}

\begin{figure}
\begin{center}
\includegraphics[width=1\textwidth,keepaspectratio]{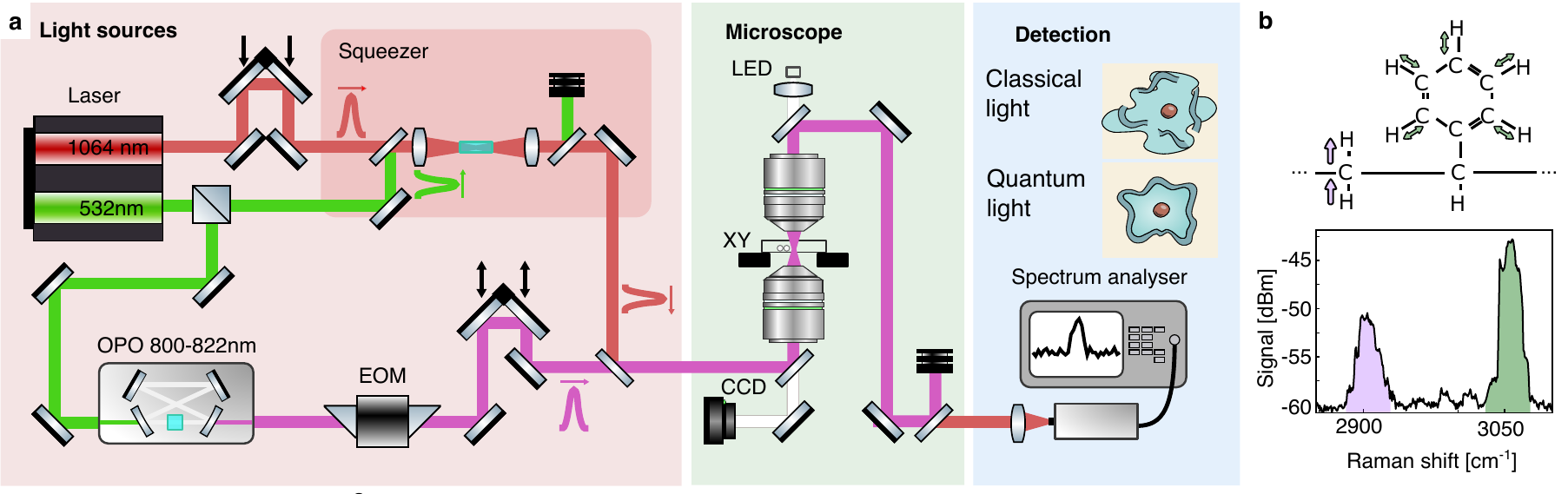} %
\caption{\textbf{Experimental setup.} 
\textbf{a,} Setup schematics. (Red panel) Preparation of the pump beam (purple) via an Optical Parametric Oscillator (OPO) and 20 MHz modulation from an Electro-Optic Modulator (EOM), and Stokes beam (red) which is amplitude squeezed in a periodically poled KTiOPO$_4$ crystal pumped with 532 nm light. (Green panel) Stimulated Raman scattering is generated in samples at the microscope focus, with raster imaging performed by scanning the sample through the focus. A CCD camera and a light emitting diode (LED) allow simultaneous bright field microscopy. After filtering out the pump, the Stokes beam is detected and the signal processed using a spectrum analyser (blue panel). \textbf{b,} Raman spectra measured in a 3~$\mu$m polystyrene bead, showing the CH$_2$ antisymmetric stretch (purple) and CH aromatic stretch (green) resonances. Taken with 100~kHz spectrum analyser resolution bandwidth (RBW).}
\label{fig:setup}
\end{center}
\end{figure}

In Raman scattering, a pump photon inelastically scatters from a molecule, exciting a chemical bond vibration and re-emitting a lower frequency Stokes photon. The frequency shift, or {\it Raman shift}, between pump and Stokes photons corresponds to the vibrational frequency of the bond, providing spectroscopic information about the molecule. However, Raman scattering is inherently weak \cite{camp2015chemically,wei2017super}. Coherent Raman microscopes enhance the process using resonant driving from multiple lasers~\cite{camp2015chemically}. The particular coherent Raman microscope constructed here is a 
stimulated Raman microscope which uses excitation lasers at 
both pump and Stokes frequencies~\cite{hu2019biological}. The Stokes laser acts to stimulate the Raman process, enhancing the scattering of pump photons.

A schematic of our custom-designed stimulated Raman microscope is shown in Fig.~\ref{fig:setup}{\bf a} (details in Supplementary Materials Sections 1-4). The Raman scattering rate, and hence signal-to-noise, depends on the product of pump and Stokes laser intensities. Consequently, we use picosecond-pulsed Stokes and pump lasers to reach high peak intensities. Near-infrared wavelengths are chosen to minimise laser absorption and photodamage in biological specimens~\cite{fu2006characterization,taylor2016quantum}. To avoid degradation of quantum correlations we employ custom high numerical aperture water immersion microscope objectives that maintain Stokes transmission $>$92\%. They are also designed to ensure tight focussing of the laser fields, and therefore high intensities and spatial resolution.
Compared to typical high quality objectives with $\sim 65$\% efficiency, the high efficiency of our objectives also increases   the number of collected Raman scattered photons by 42\%, with a commensurate increase in signal strength. At the output of the microscope, the Stokes light is detected on a custom-designed  photodetector with very low electronic noise and high bandwidth. Together with previously established laser noise minimisation techniques that shift the Raman signal into modulation sidebands around the Stokes frequency~\cite{Xie2008}, this allows shot-noise limited operation with relative intensity noise comparable to state-of-the-art stimulated Raman microscopes~\cite{Xie2008,saar2010video} (Supplementary Materials Section 3).

\section*{Photodamage constrains microscope performance}

Before demonstrating that quantum correlations can be used to evade photodamage in 
 Raman microscopy, we systematically explore the shot-noise limited performance of the microscope using dry monodisperse samples of polystyrene beads.
Fig.~\ref{fig:setup}{\bf b} shows a typical Raman spectrum,  exhibiting several Raman bands. The two strongest bands are the CH$_2$ antisymmetric stretch 
and the CH aromatic stretch,
illustrated in the figure inset. We focus here on the latter.

\begin{figure}
\begin{center}
\includegraphics[width=1\textwidth,keepaspectratio]{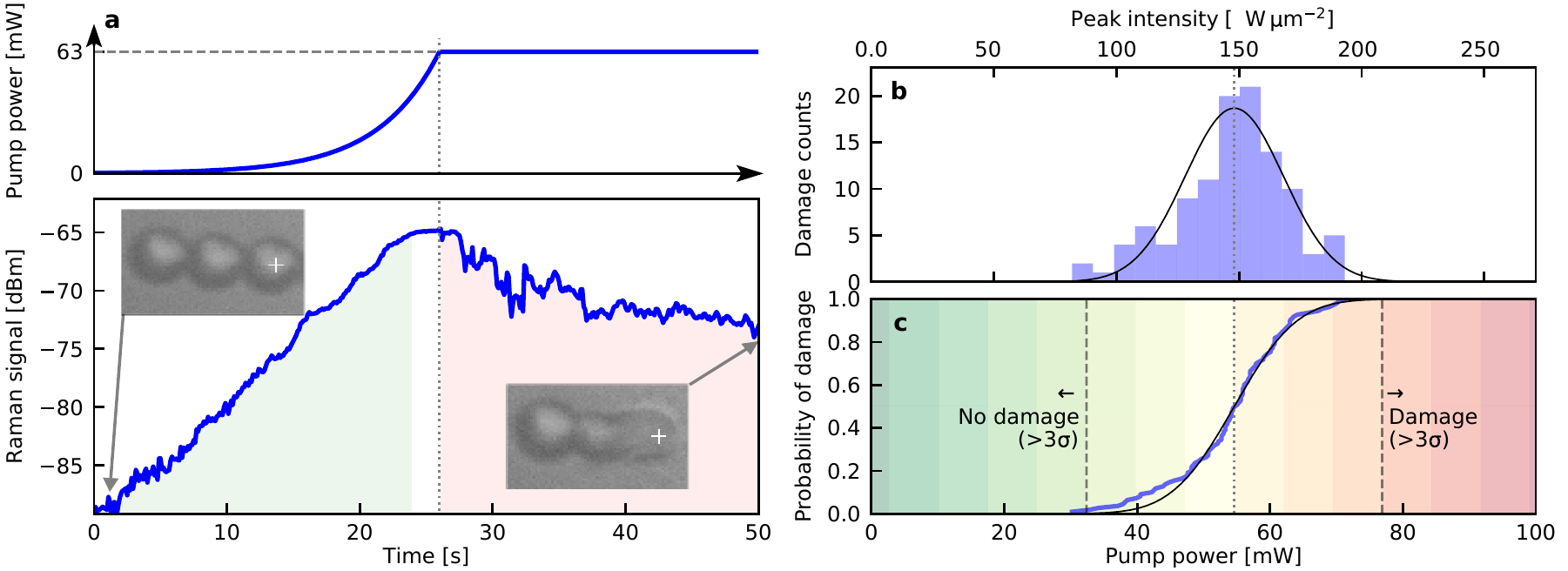} 
\caption{
\textbf{Quantifying photodamage.}
The Raman signal from a 3~$\mu$m polystyrene bead is probed as a function of pump power with a fixed 3~mW Stokes power.
\textbf{a,} {\it top}: The pump power is gradually increased until photodamage is observed, in this example occurring at a power at the sample of 63~mW. It is then held fixed. {\it Bottom}: The Raman signal increases with power prior to photodamage (green) and drops at fixed pump intensity after the particle is damaged (red). Visual inspection (insets) confirms that photodamage has occurred. RBW: 1~kHz. \textbf{b,c} Characterization of the photodamage threshold for 110 particles. \textbf{b,} Histogram of the pump power at which damage was observed, with Gaussian fit  of the histogram showing the mean photodamage threshold at 54.6 mW (dashed line), corresponding to a peak intensity of 150 W/$\mu$m$^2$. \textbf{c,} Cumulative distribution of the power at which photodamage occurs, with dotted line indicating the mean and each shading bar corresponding to +1$\sigma$ from the mean.}
\label{fig:damage}
\end{center}
\end{figure}

To characterise possible photodamage induced by the pump and Stokes lasers
we monitor the Raman signal from the CH aromatic stretch band as the pump power is gradually increased over time. At low pump powers, the signal increases as expected. However, at high powers the signal diverges, typically decreasing markedly and decaying over time as shown for example in Fig.~\ref{fig:damage}{\bf a}. This is a signature of the onset of photodamage. We verify this using bright field imaging which, for this example measurement, shows substantial deformation of the polystyrene bead due to absorptive heating,  as well as deformation of an adjacent bead (see Fig.~\ref{fig:damage}{\bf a}). Measurements on other beads show that other damage modalities also occur, including photoablation due to the high peak pulse intensities (see Supplementary Materials Sections 5\&6). The observation of both mean and peak intensity dependent photodamage modalities is an indication that the laser repetition rate and pulse length are well chosen to balance the different classes of photodamage~\cite{fu2006characterization}.

To determine the photodamage statistics, we repeat the measurement on 110  beads, and record the threshold pump intensity at which the observed signal visibly diverges from damage-free expectations. Fig.~\ref{fig:damage}{\bf b} plots the distribution, which fits well to a  Gaussian with mean power at the sample of 54.6~mW and standard deviation of  7.4~mW. The mean power corresponds to a peak pulse intensity of around 150~W/$\mu$m$^2$ consistent with known thresholds for laser ablation~\cite{Oraevsky1996}. 
The probability of photodamage is highly sensitive to the optical intensity, as shown  by the cumulative probability distribution in Fig.~\ref{fig:damage}{\bf c}.
For instance, when using a pump power of 30~mW none of the polystyrene beads exhibited photodamage, while 26\% were damaged at 50~mW.

\section*{Quantum-correlations allow absolute performance advantage}

\begin{figure}
\begin{center}
\includegraphics[width=1\textwidth,keepaspectratio]{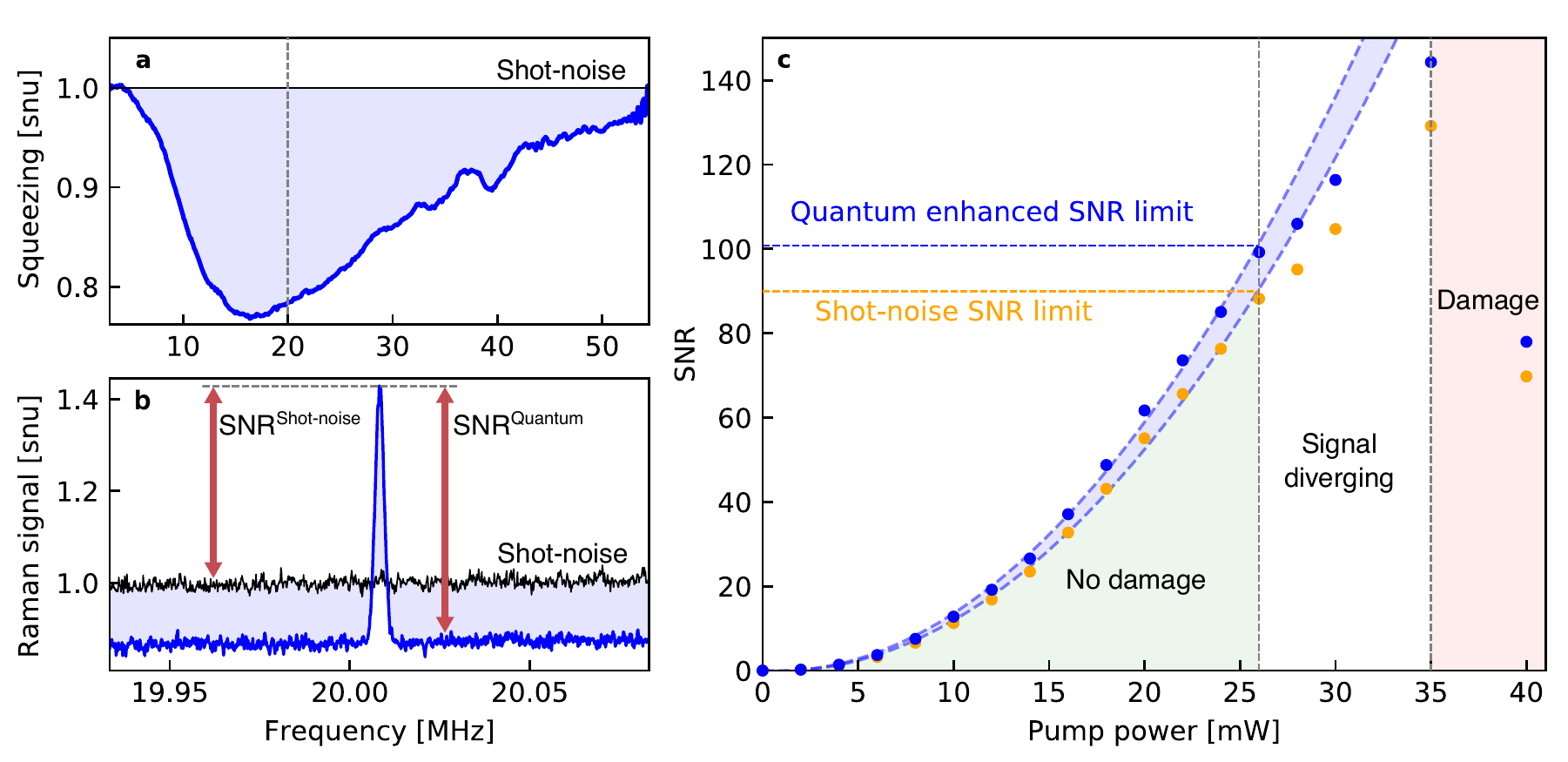} 
\caption{\textbf{Quantum-enhanced stimulated Raman microscopy.}
\textbf{a,} Squeezed light spectrum  
normalised to shot-noise (snu: shot-noise units). The maximum squeezing is near the 20 MHz Raman modulation (vertical dashed line). RBW: 1~MHz. \textbf{b,} Stimulated Raman signal of a 3~$\mu$m polystyrene bead using 3~mW of pump light at the sample. Squeezed light reduces the total measurement noise to 13\% below shot-noise (or $-0.60$~dB), improving the signal-to-noise ratio (SNR) by 15\%. \textbf{c,} SNR for one 3~$\mu$m polystyrene particle with increasing pump power. The quantum-enhanced SNR is determined directly, with the shot-noise-limited SNR inferred from the ratio of shot-noise-limited  and quantum-enhanced noise floors. This ensures a fair comparison, insensitive to spatiotemporal modeshape variations between the measurements (see Supplementary Materials Section 8.3). The statistically-determined one-sigma uncertainty in all SNRs is less than 1\%.
The SNR increases quadratically as expected until 26~mW (green), indicating the particle is undamaged.  The common fluctuations of the shot-noise-limited and squeezed SNRs about this quadratic dependence are consistent with 1\%-level laser intensity drift.
 Above 26~mW the SNR rises more slowly (white), suggestive of some disruption within the particle, and finally drops (red) when the power reaches the threshold for visible damage. Dashed lines: fits to undamaged SNR; blue shading: quantum enhancement.
  {\bf b}\&{\bf c,} RBW: 3~kHz.}
\label{fig:quantum}
\end{center}
\end{figure}

The concentration sensitivity of a Raman microscope is directly related to its signal-to-noise (Supplementary Materials Section 11).
To increase the signal-to-noise 
beyond the constraint due to photodamage, we introduce quantum correlations between Stokes photons. The quantum correlations are generated using a home-built optical parametric amplifier that produces bright amplitude squeezed light (see Supplementary Materials Sections 7-10). This allows the measurement noise floor to be reduced beneath the shot-noise. Fig.~\ref{fig:quantum}{\bf a} shows the resulting reduction in noise variance as a function of frequency, with 22\% (or -1.1~dB) reduction achieved here at the frequency of the stimulated Raman modulation sidebands.  
Similar quantum correlations -- albeit continuous-wave with far lower peak intensity -- have been applied in several biophysically-relevant proof-of-principle experiments, demonstrating improvements in sensitivity (e.g.~\cite{pooser2016plasmonic,dowran2018quantum, taylor2013biological}), resolution~(e.g.~\cite{taylor2014subdiffraction, tenne2019super, israel2014supersensitive}) and fundamental capabilities~\cite{phan2014interaction, kalashnikov2016infrared, paterova2018tunable}. Used here they allow Raman scattering  to be observed even when less than one photon is scattered on average during the measurement interval, removing what is otherwise a strict constraint on photon budget~\cite{camp2015chemically,schermelleh2019super, hoover2013advances}.

To demonstrate quantum-enhanced performance we inject the squeezed Stokes laser into the stimulated Raman microscope while probing the CH aromatic stretch band of a 3~$\mu$m polystyrene bead. We then measure the power spectrum of the detected photocurrent. A typical spectrum is shown in Fig.~\ref{fig:quantum}{\bf b}. The peak in the spectrum is the stimulated Raman signal. Losses on transmission through the microscope, pump filters, sample coverslips and polystyrene bead of around 35\% degrade the quantum correlations. Nevertheless, the noise floor is still reduced by a factor of 13\%, consistent with expectations given the additional loss (Supplementary Materials Section 10).

The signal-to-noise
 is defined as the ratio of the height of the stimulated Raman peak to the power spectrum noise floor. We find that it
initially increases quadratically with Raman pump power,
as expected for stimulated Raman scattering~\cite{Xie2008}. However, as shown for one example in Fig.~\ref{fig:quantum}{\bf c}, the manifestation of photodamage prevents this scaling from continuing indefinitely. When using shot-noise limited light there is a strict maximum in damage-free signal-to-noise, which for the example of Fig.~\ref{fig:quantum}{\bf c} is SNR$_{\rm max}^{\rm shot-noise} = 88.2 \pm 0.2$ from a measurement with 3~kHz resolution bandwidth. Quantum correlations provide a damage-free signal-to-noise of SNR$_{\rm max}^{\rm quantum} = 99.2 \pm 0.3$, corresponding to an enhancement of SNR$_{\rm max}^{\rm quantum}/$SNR$_{\rm max}^{\rm shot-noise}-1 = 12.5 \pm 0.4$\%. 
This represents an absolute quantum advantage --- with same spatiotemporal modeshapes and apparatus, photodamage prohibits classical techniques from reaching this level of signal-to-noise. It contrasts previous demonstrations of quantum-illumination enhanced imaging~\cite{brida2010experimental, defienne2019quantum, sabines2019twin, samantaray2017realization,israel2014supersensitive, ono2013entanglement}, where equivalent performance could be attained by increasing the classical illumination intensity.

 We note two parallel demonstrations of  quantum-enhanced nonlinear spectroscopy~\cite{andrade2020quantum,triginer2020quantum}. Both use peak optical intensities well below the levels used in state-of-the-art nonlinear microscopes, and do not observe photodamage. Nor do they perform imaging, though Ref.~\cite{andrade2020quantum} does  report spatially distributed measurements.
Our combination of pulsed excitation and high numerical aperture objectives enables orders-of-magnitude
higher peak intensities for both pump and Stokes. This brings the microscope into the regime where photodamage is relevant, and increases the nonlinear signal amplitude (proportional to the product of peak intensities) by around a factor of $10^8$.

\section*{Quantum-enhanced imaging}

\begin{figure}
\begin{center}
\includegraphics[width=1\textwidth,keepaspectratio]{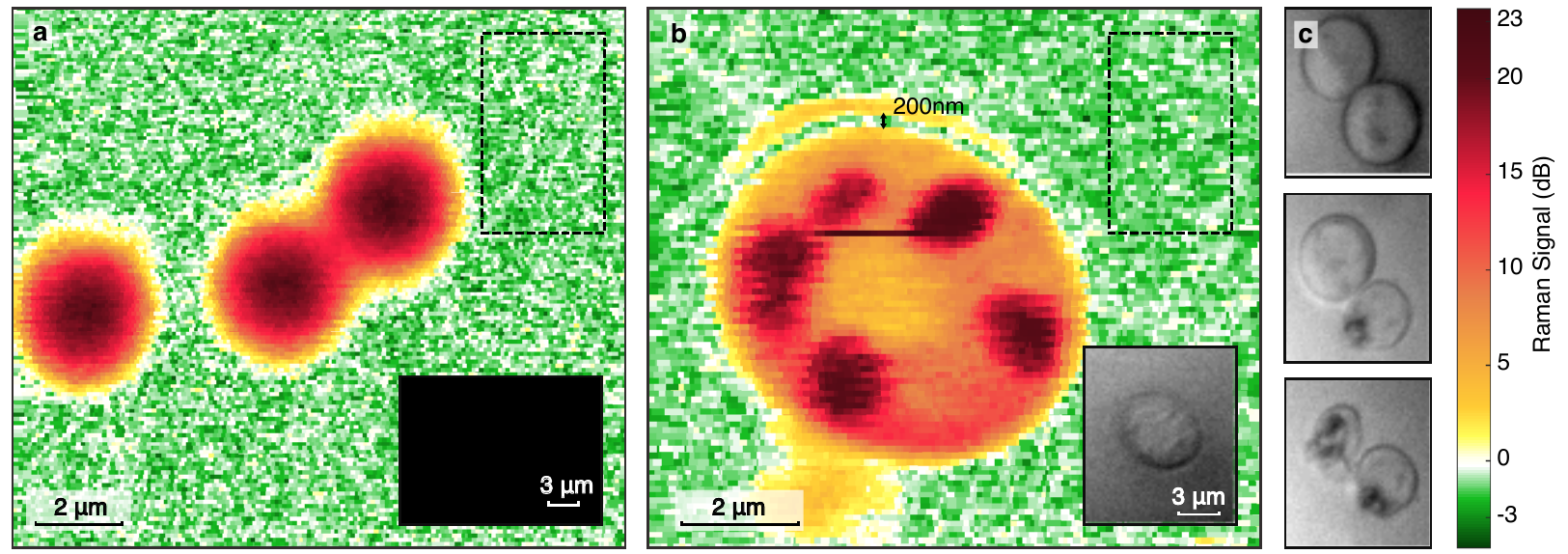} 
\caption{\textbf{Quantum enhanced imaging.}  
 {\textbf a} Image of 3~$\mu$m polystyrene beads at a Raman shift of 3055~cm$^{-1}$ obtained with 6~mW of pump power at the sample. The background (coloured green) has no Raman signal, and is limited by measurement noise which is 0.9~dB below shot-noise, providing a 23\% increase in SNR.
{\textbf b}, Image of a live yeast cell ({\it Saccharomyces cerevisiae}) in aqueous buffer at 2850~cm$^{-1}$ Raman shift.   Several organelles are clearly visible. The faint outline of what may be the cell membrane is also visible, showing that the microscope has a resolution of around 200~nm.
Here, the measurement noise is reduced by 1.3~dB below shot-noise, corresponding to a 35\% SNR improvement.
This image was recorded with  $\sim$30~mW of pump power at the sample. The pump intensity of 210~W~$\mu$m$^{-2}$ was beneath  that at which visible cell damage was observed. Dashed rectangular boxes in {\bf a}\&{\bf b} show the regions used to determine the measurement noise, and insets are bright-field microscopy images. {\bf c,} A sequence of images in which two cells are illuminated with the same pump power as in {\bf b} but focussed to roughly a factor of two higher intensity, producing visible photodamage after less than a minute of exposure. 
For both {\textbf a}\&{\textbf b}, RBW: 1~kHz. 
}
\label{fig4}
\end{center}
\end{figure}

To demonstrate quantum-enhanced imaging, we record the power of the stimulated Raman signal as the microscope sample stage is raster scanned over samples of both dry polystyrene beads and living {\it Saccharomyces cerevisiae} yeast cells in aqueous solution (details in Supplementary Materials Section 12). A relatively long pixel dwell time of 50~ms is used, limited by our stage scanning system. 
The 40~MHz bandwidth  of quantum enhancement demonstrated  (Fig.~\ref{fig:quantum}{\bf a}) is compatible with the faster scanning systems necessary for video rate imaging~\cite{saar2010video},
indicating that quantum enhanced video-rate imaging should be possible in future.

Fig.~\ref{fig4}{\bf a} shows a typical quantum-enhanced image of a collection of 3~$\mu$m polystyrene beads, with signal-to-noise enhanced by 23\% compared to the shot-noise limit. 
Fig.~\ref{fig4}{\bf b} shows the equivalent image for a single yeast cell, in this case recorded at a Raman shift of 2850~cm$^{-1}$ to target the CH$_2$ bonds that are most prevalent in lipids. 
 Improved alignment of the microscope and squeezed light source, together with lower Fresnel reflective losses at water-glass interfaces compared to air, in this case allow a 35\% enhancement in signal-to-noise, increasing 
  the contrast of the image (see direct comparison in Supplementary Materials Section 12.2). Visible cell damage was observed at higher pump intensities (Fig.~\ref{fig4}{\bf b} {\it right}). Without quantum correlations or exposing the sample to these higher intensities, a 35\% higher pixel dwell time would be required to achieve the same contrast, which would reduce the frame rate of the microscope. The enhanced contrast is particularly useful in subcellular imaging since many features have far-sub-wavelength dimensions and produce a correspondingly small Raman signal. 

When imaging yeast cells, the microscope benefits fully from the use of high numerical aperture water immersion objectives which, together with the nonlinearity of the Raman interaction, enable sub-wavelength resolution (Fig.~\ref{fig4}{\bf b}). Compared to the bright field microscope image in the inset of Fig.~\ref{fig4}{\bf b}, the stimulated Raman images provide significantly more information about the interior cellular structure. Raman signal is visible across the entire cell volume, including five bright organelles which have size and composition consistent with lipid droplets~\cite{kochan2018single}. A thin semi-circular feature is also resolved that is consistent with a section of the cell membrane. Raman signals from the cell membrane are typically faint due to its 10~nm thickness.  Here, quantum-correlations allow an approximately 40\% longer length of membrane to be resolved than would have been possible with coherent illumination (see Supplementary Materials Section 12.2).

\section*{Discussion and outlook} \label{Sec:Discussion}

Photodamage-evading microscopes are broadly recognised as a key metrological application of quantum technologies, proposed some three decades ago~\cite{Slusher}. For instance, they are one of two ten-year quantum-enhanced imaging milestones in the UK Quantum Technologies Roadmap~\cite{UKQTroadmap}.
Our work realises such a microscope.
We demonstrate that quantum correlations allow performance that could not be achieved by simply increasing illumination intensity, and therefore that they afford an absolute quantum advantage. 
This  provides a path to exceed severe constraints on existing high performance microscopes
that would otherwise be fundamental~\cite{li2020adaptive, schermelleh2019super},  allowing their contrast to be improved at fixed frame rate or increased frame rates without exposure to additional risk of photodamage.
We further show
 that quantum-correlated illumination can enhance imaging of the interior of a living cell. Our implementation within a coherent Raman microscope provides the capacity for wide impact due to the extremely high specificity and label-free operation such microscopes provide. Coherent Raman microscopes have seen broad applications over the past decade~(e.g. see \cite{camp2015chemically,cheng2015vibrational,hu2019biological}). With this progress, both sensitivity and speed are now limited by the constraint that photodamage places on optical intensities. Faster and more sensitive imaging currently requires alternative methods such as fluorescence imaging, for which labels provide far higher cross-sections than are available in label-free Raman scattering. This is a barrier to important applications such as video-rate imaging of weak molecular vibrations and label-free spectrally resolved imaging~\cite{wei2017super, cheng2015vibrational},  a barrier that our approach provides the means to overcomes. 

In absolute terms, the level of improvement achieved here is relatively low. This is due, in large part, to the relative immaturity of technology capable of generating and detecting bright picosecond squeezed light, together with the relatively low total optical efficiency for squeezed light in our apparatus of approximately 40\%. While it may be challenging to achieve high collection efficiency {\it in vivo}~\cite{saar2010video};  absorption and scattering of {\it in vitro} samples is typically very low, as evidenced by the poor contrast of simple bright-field imaging,
and therefore should not present a significant constraint. Continuous wave squeezing technology that provides as much as a factor of thirty enhancement in signal-to-noise has been developed  
to improve the capabilities of gravitational wave detectors \cite{vahlbruch2016detection}. Even allowing for additional optical loss from two high numerical aperture objectives, this suggests that an order-of-magnitude improvement in damage-free signal-to-noise is feasible in future quantum-enhanced coherent Raman microscopes, or equivalently that their frame rate could be increased by the same margin without reducing signal-to-noise. Such an improvement would only otherwise be possible using contrast agents with large Raman scattering cross-sections~\cite{wei2017super}, sacrificing label-free operation. Quantum correlations could also be used to operate with lower light intensities and thereby suppress performance-limiting noise processes such as out-of-focus fluorescence 
and background scatter~\cite{hoover2013advances}, to improve the single-molecule imaging sensitivity of plasmon-enhanced Raman microscopes~\cite{zong2019plasmon}, or even to enhance the cross-section of Raman scattering  and therefore signal strength~\cite{michael2019squeezing}.

\section*{References}
 
 \bibliographystyle{unsrt} %to change the citation number in the text automatically
 
 \bibliography{mybib}
 
\section*{Data availability} 

Supplementary Information is available for this paper. Further data that supports the findings of this study are available from the corresponding author upon reasonable request.

\section*{Acknowledgements} 

We acknowledge Walter Wasserman for sourcing the yeast cells in trying circumstances, Ulrich Hoff for contributions to the construction of the apparatus, and APE GmbH for support related to the laser system. This work was supported primarily by the Air Force Office of Scientific Research (AFOSR) grant FA2386-14-1-4046. It was also supported by the Australian Research Council Centre of Excellence for Engineered Quantum Systems (EQUS, CE170100009). W.P.B. acknowledges the Australian Research Council Future Fellowship, FT140100650. M.A.T. acknowledges the Australian Research Council Discovery Early Career Research Award,   DE190100641.

\newpage

\part*{Supplementary Materials}

\newcounter{eqnctr}
\renewcommand{\thefigure}{S\arabic{figure}}
%\title{Supplementary Information}

\setcounter{section}{0}
\setcounter{figure}{0}

This Supplementary Material provides details about the theory, modelling, and experimental techniques used in the paper ``\emph{Quantum correlations overcome the photodamage limits of light microscopy}''.

\section{Microscope design}

The microscope is home-made to allow both conventional bright-field imaging and quantum-enhanced stimulated Raman imaging (see Fig.~1{\bf a} in the main text). The microscope frame is home-built in an inverted configuration. Light is both focused and collected in the microscope using two matching water immersion objectives which were custom made by Leica Microsystems. These objectives have custom anti-reflection coating at 1064nm applied in Leica HC-PL-APO63x/1.20W objectives, which ensures ultrahigh Stokes transmission of $>$92\%. To ensure this high transmission is maintained in both dry and aqueous media we under-fill the objectives (see Section~\ref{est_I}). The objectives have a relatively low $\sim$32\% transmission efficiency for pump light. However, the pump intensity is constrained by the photodamage it causes to the sample rather than the laser source (see Fig.~2 of the main text), and is not sent to the detector (see Fig.~1{\bf a} of the main text) so this attenuation of the pump has no consequence for the achievable signal-to-noise.  

The objective focus is controlled using a Mad City Labs Nano-F200S focusing module. Samples are positioned with nanometer scale precision using a Mad City Labs Nano-Bio100 stage, which was used to perform raster scanning in imaging experiments. Samples are simultaneously illuminated with incoherent light from a green LED and imaged using a camera (Allied Vision Manta).

\section{Laser sources for stimulated Raman excitation}

The Stokes laser is provided by a low noise solid-state pulsed laser with  6~ps pulse length, an 80~MHz repetition rate, and a wavelength of $\lambda_{\rm Stokes}=1064$~nm (Coherent, Plecter Duo). Part of the laser output is frequency-doubled to 532~nm and used to pump both the home-built squeezer (described below) and a commercial optical parametric oscillator (OPO; APE, Levante Emerald). The OPO generates a 500 to 700~mW pump laser that is temporally-synchronised and frequency-tuneable over a wavelength range from 750 to 900~nm.  This provides the capacity to probe Raman shifts over the range 1713--3935~cm$^{-1}$. 

The pulse duration of the laser source was chosen to be comparable to the decay time of molecular vibrations since this maximises the Raman interaction, with molecular decay times typically in the range of a few picoseconds~\cite{Xie2008}. The choice of 80~MHz repetition rate is designed to approximately minimise photodamage processes~\cite{fu2006characterization}. Lower repetition rates lead to higher peak intensity for a given average intensity, thereby increasing the relative contribution of nonlinear damage processes like ablation, while higher repetition rate increases the relative contribution of linear damage for instance due to heating. 

Quantum correlations are generated between photons in the Stokes laser by passing the beam through a home-built squeezed light source, described below.

\section{Reaching the shot-noise limit}

Many factors can prevent precision optics experiments from operating at the shot-noise floor; most particularly technical laser noise and photodetector electronic noise. Both of these noise sources are especially problematic at low frequencies. To evade low frequency noise, we implement the technique developed in Ref.~\cite{Xie2008} wherein the pump laser is strongly amplitude modulated.
This modulates the stimulated Raman process, shifting the detected signal to sidebands around the modulation frequency, and away from the low frequency laser and electronic noise. In our case we use a resonant electro-optic modulator that provides a high depth of modulation at 20.0083~MHz (approximated as 20~MHz elsewhere in the paper). A custom-built photodetector  allows the detection of the sidebands with noise floor dominated by optical shot-noise (or the reduced squeezed noise) and good clearance from both technical laser noise and photodetector electronic noise.

Using this apparatus and approach, we find that the detected Stokes light is shot-noise limited (see Section~\ref{verif}). It is then possible to estimate the relative intensity noise $\Delta I / I $ of our stimulated Raman microscope without squeezing. We find  $\Delta I / I = (2 \hbar c / \eta P \lambda_{\rm Stokes}  \tau)^{1/2} \sim 10^{-8}$, competitive with the best state-of-the-art stimulated Raman microscopes~\cite{Xie2008,saar2010video}, where $P=$3~mW is the Stokes power reaching the detector, $c$ the speed of light, $\tau=1$~s is the integration time, and $\eta=72$\% the measured quantum efficiency of the photodiode. Using squeezed light further reduces this noise.

\section{Stimulated Raman scattering}

To detect and characterise the stimulated Raman scattering, the Stokes laser is measured on a resonant feedback photodetector (see Section~\ref{det_des}) and processed with a spectrum analyser. The average Stokes power reaching the detector is kept constant at 3~mW for all experiments, ensuring both shot-noise limited (or squeezed when the optical parametric amplifier is on) performance and large clearance from the electronic noise of the detector. As discussed above, the stimulated Raman process causes the 20~MHz modulation of the pump field to be imprinted on the Stokes field, with the size of the imprinted modulation proportional to the strength of Raman scattering. As such, measurement of the 20~MHz modulation of the Stokes field provides a direct measurement of the Raman gain at the frequency shift between the pump and Stokes fields. The Raman spectrum (Fig.~1{\bf b} of the main text) was measured in polystyrene beads by monitoring this 20~MHz signal while adjusting the OPO to sweep the pump laser wavelength between 800 and 816~nm. 

\section{Sample preparation}

The Raman signal is characterized using polystyrene microspheres. Polystyrene has a relatively simple spectrum with strong Raman bands with Raman shifts in the range of 600 to 3100~cm$^{-1}$, accessible using our pump and Stokes lasers. The uniform composition and size of the beads allows this Raman spectrum, and the photodamage induced by the measurement, to be accurately and reproducibly characterised. It also allows accurate benchmarking of the performance of the microscope.

Polystyrene bead samples were prepared from dry powder of monodisperse spherical polystyrene beads with $3~\mu m$ diameter. These were scattered onto a glass coverslip. Vacuum grease was spread around the edge of the coverslip as a sealant, and a second coverslip pressed on top.

For cellular imaging, dehydrated yeast cells ({\it Saccharomyces cerevisiae}) were sourced from a local supermarket, and hydrated in distilled water mixed with sugar (approximately 0.1$\%$ by weight) and NaCl salt (approximately 0.1$\%$) to provide sustenance and prevent osmotic shock. The cells were kept in this media for at least an hour to provide time to recover from dehydration. Cells were then transferred into sample chambers by pipetting 15~$\mu$L of the cell solution onto a glass coverslip, adding a sealant of vacuum grease around the outer edge of the coverslip, and then pressing a second coverslip on top. The grease forms a seal around the edge while the cell solution fills the approximately 70~$\mu$m gap between the coverslips. The sample was then left to rest for an hour before experiments to provide time for the cells to adhere to the surface, which is necessary for raster imaging. The health of the cell culture was verified separately by adding extra sugar and observing the production of carbon dioxide.

\section{Quantifying optical damage}

The damage threshold was characterised for samples of $3~\mu m$ polystyrene beads. To find the damage threshold, stimulated Raman signals were recorded in beads as the pump power was gradually increased. The Raman signal generally increases with pump power below the damage threshold, but decays after the onset of damage  (with an example shown in Fig.~2{\bf a} of the main text). In all cases the damage was confirmed to have occurred by bright-field imaging. This protocol was repeated for 110 particles with the histogram of recorded damage thresholds shown in  Fig.~2{\bf b} of the main text.

\begin{figure}
\begin{center}
\includegraphics[width=0.5\textwidth,keepaspectratio]{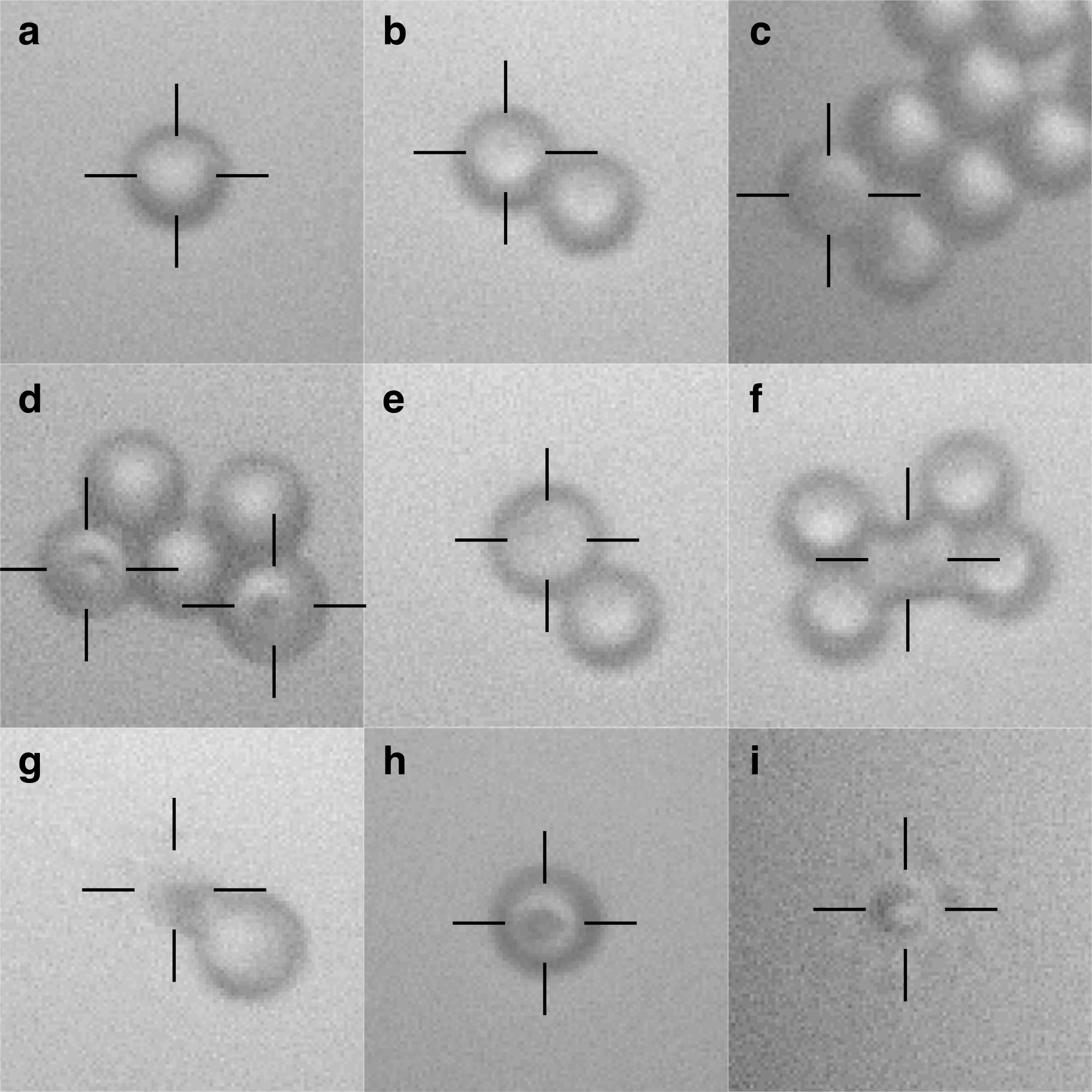} 
\caption{\textbf{Bright field images of typical photodamage.}  
 Examples of the different photodamage modalities that were observed for $3~\mu$m polystyrene beads. {\bf a,} Undamaged bead.  {\bf b,} Slight deformation or ablation.  {\bf c,} Deformation of target bead and adjacent bead.  {\bf d,} Ablation of two beads that were separately exposed.  {\bf e,} Deformation.  {\bf f,} Significant deformation.  {\bf g,} Complete destruction.  {\bf h,} Ablation.  {\bf i,} Complete disintegration. In each image, the cross-hairs indicate the target bead(s).
}
\label{zoo}
\end{center}
\end{figure}

Several different damage modalities were observed, as illustrated by the example bright field microscope images in Fig.~\ref{zoo}, including deformation due to photoabsorptive heating, ablation due to the high peak intensity of the pump laser, and complete destruction/disintegration.

\subsection{Estimating the peak pump intensity at the sample}
\label{est_I}

To estimate the peak pump intensity $I_{\rm peak}$ at the sample from the average pump power at the sample $P$ we assume that the pump pulses have a Gaussian temporal profile with full-width-half-maximum $t_{\rm FWHM}=6$~ps, along with a Gaussian transverse profile with width at the focus $w_0$. Given the second of assumptions, it is straightforward to show that the peak pump intensity is given by
\begin{equation}
I_{\rm peak} = \frac{P_{\rm peak}}{(\pi w_0^2/2)},
\end{equation}
where $P_{\rm peak}$ is the peak pump power at the sample. The first assumption allows us to relate the peak power to the average power as
\begin{equation}
P_{\rm peak} = 2 \sqrt{\frac{\ln 2}{\pi}} \frac{\tau}{t_{\rm FWHM}} P,
\end{equation}
where here $\tau = 1/({\rm pulse~repetition~rate~in~MHz}) = 12.5$~ns is the repetition time of the pump pulse train.

The waist size can be approximated as
\begin{equation}
w_0 = \frac{\lambda_{\rm pump}}{\pi {\rm NA_{effective}}},
\end{equation}
where $\lambda_{\rm pump}=1064$~nm is the pump wavelength in vacuum and ${\rm NA_{effective}}$ is the filled numerical aperture of the objective lens. Overfilling the objective was found to introduce background noise due to nonlinear cross-phase modulation effects. All measurements on polystyrene beads were made with an effective numerical aperture of ${\rm NA_{effective}} \sim {\rm NA}/2.5 \sim 0.5$, where NA$=1.2$ is the numerical aperture of the microscope. The yeast cell images for which no damage was observed were obtained with  ${\rm NA_{effective}} \sim 2{\rm NA}/3 =0.8$. The yeast cell images showing photodamage were taken with ${\rm NA_{effective}} \sim {\rm NA}=1.2$, and therefore the intensity increased by a factor of $(3/2)^2 \sim 2$.

\section{Production of bright squeezed light}

\begin{figure}[b!]
\begin{center}
\includegraphics[width=0.7\textwidth]{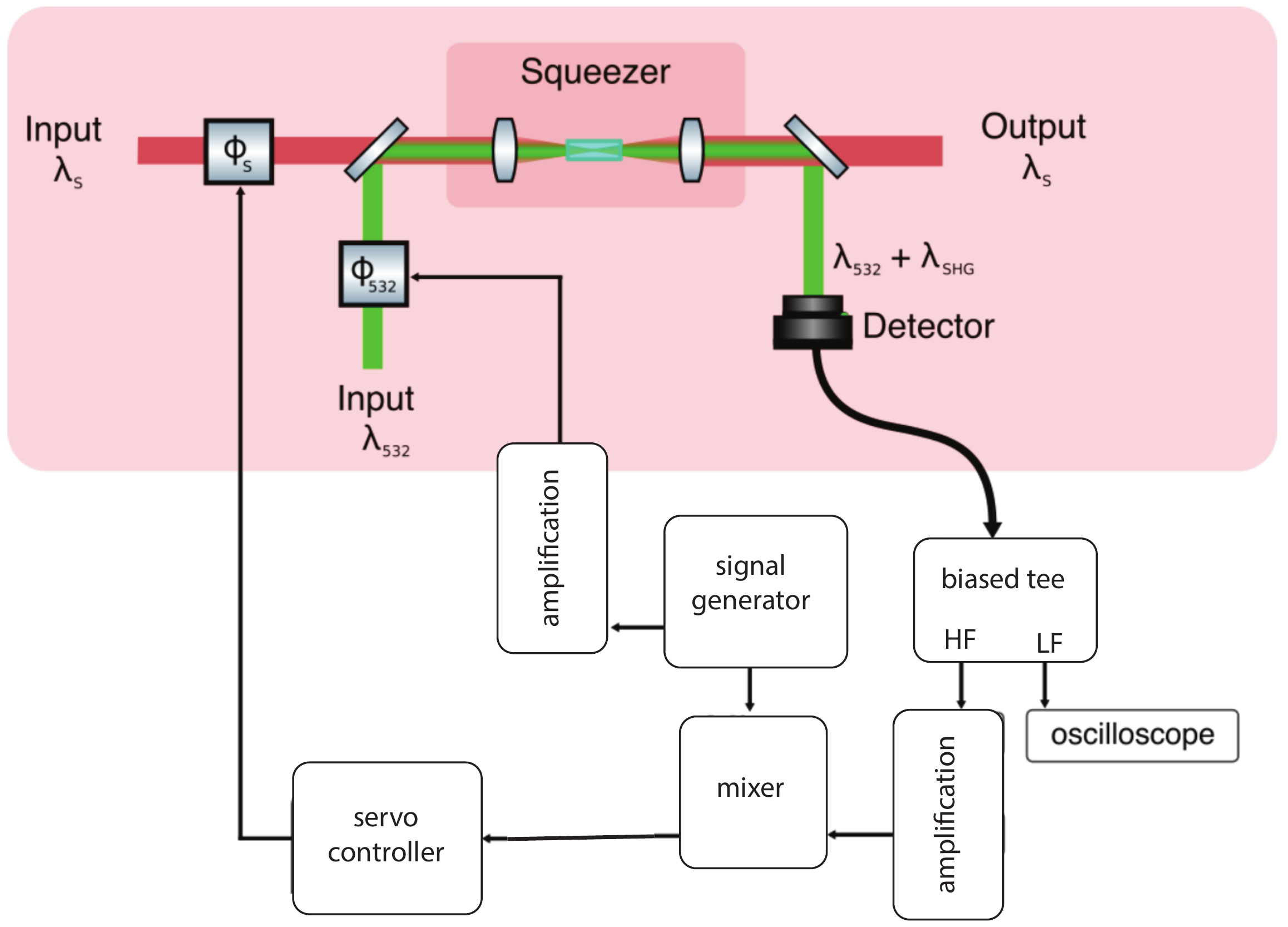}
\end{center}
\caption{\label{fig:schem} Schematic of squeezed light source. $\phi_s$: phase shift on 1064~nm Stokes beam. $\phi_{532}$ phase shift on 532~nm squeezer pump beam. $\lambda_{1064}$ and $\lambda_{532}$ wavelengths of Stokes and squeezer pump. LPF: low pass filter. HF: high frequency. LF: low frequency.
}
\end{figure}

To produce squeezed light we utilise a single pass optical parametric amplifier (OPA) that consists of a second-order nonlinear crystal which is pumped at 532~nm and seeded  by the Raman Stokes field at 1064~nm (see Fig.~\ref{fig:schem}). The OPA pump is generated by frequency doubling the 1064~nm light from the Coherent Plecter Duo laser, also used to generate the Stokes field. The nonlinear crystal is a 7~mm long Periodically Polled Potassium Titanyl Phosphate (PPKTP) crystal (Raicol Crystals, s/n:7-112714933). The crystal is temperature stabilised to 32$^\circ$C to phase match the nonlinear process. The relative phase between the seed and OPA pump is locked using a standard Pound-Drever-Hall type locking system, with a 104.7~kHz modulation applied to the green field, the green light transmitted through the nonlinear crystal detected and demodulated at 104.7~kHz, and the resulting error signal applied back to the phase of the 1064~nm Stokes field. Choosing the sign of the lock correctly allow the squeezing light source to be stably locked to deamplify the Stokes field, leading to amplitude squeezing.  The output state is a bright amplitude squeezed state of light which exhibits pair-wise correlations between photons~\cite{taylor2016quantum}. %For all experiments reported here we use 3~mW of detected Stokes light.

\section{Detecting squeezing and performing quantum-enhanced measurements}

To detect the squeezed light, and later to implement quantum-enhanced Raman microscopy, we choose to use a direct detection scheme rather than the usual homodyne detection. Generally homodyne detection is preferred since the interference with the local oscillator both acts to boost the shot-noise of the signal beam above the electronic noise of the detector, and because it provides a straightforward and robust way of determining the shot-noise level, and therefore the level of quantum enhancement\footnote{By blocking the signal beam, so that the homodyne detects the noise level of the vacuum entering its signal port.}.

The primary reason to choose direct detection here is that homodyne detection generally requires a local power at least an order of magnitude higher than the signal power. Unlike most experiments with squeezed light, in our experiments the signal power must be relatively high to reach parameter regimes relevant to state-of-the-art Raman microscopy and for which photodamage starts to become apparent.  The local oscillator powers required for homodyne detection would then introduce significant challenges in detector design. A second reason to choose direct detection is that it avoids the need to mode-match the signal field to another field after it has passed through the microscope, and therefore relaxes requirements on the beam quality of the transmitted signal field.

\begin{figure}[]
\begin{center}
\includegraphics[width=1\textwidth]{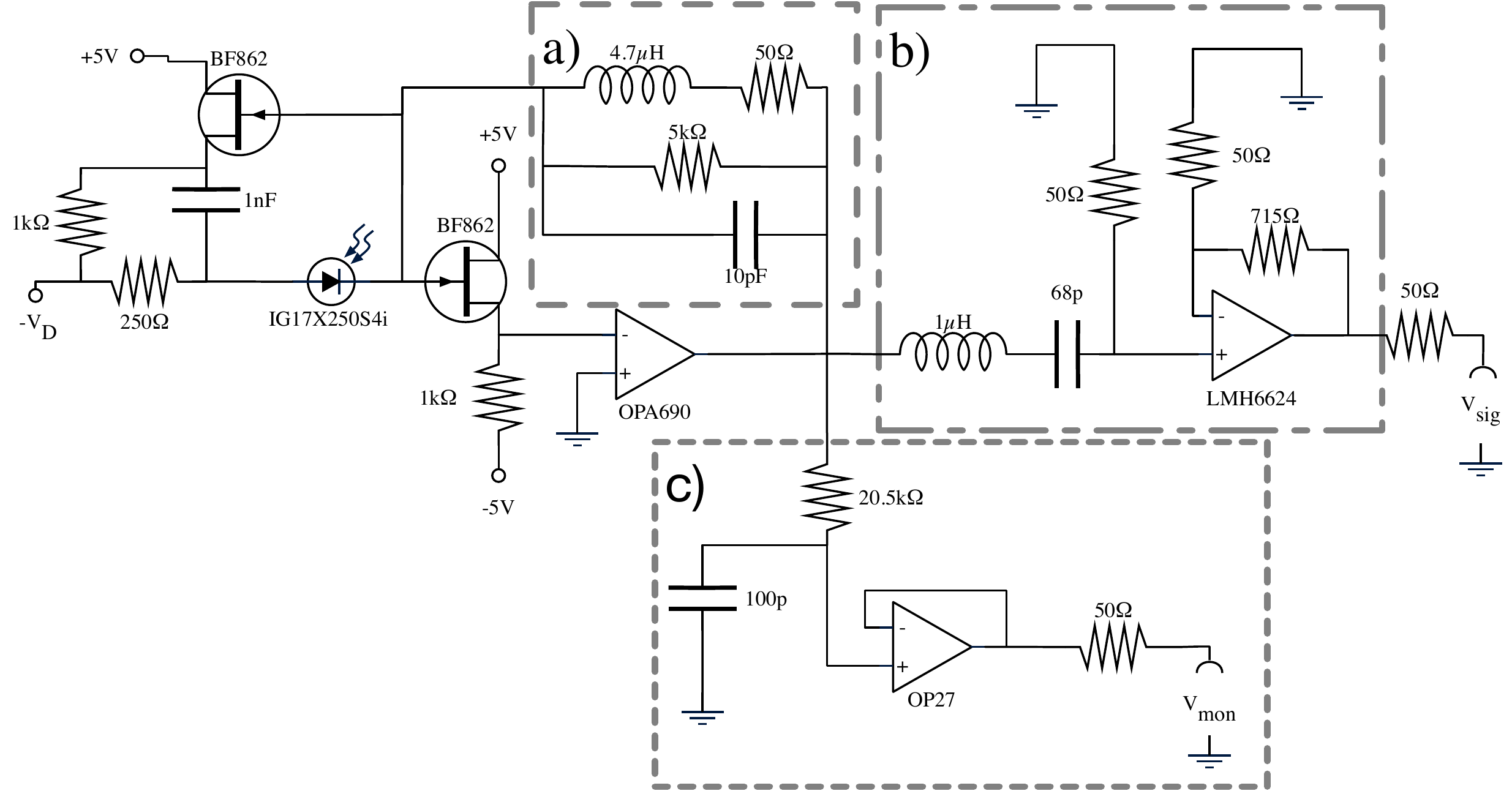}
\end{center}
\caption{\label{fig:detector}  Circuit diagram for the custom-built detector used to detect squeezed light and perform quantum-enhanced Raman microscopy.
}
\end{figure}

\subsection{Detector design}
\label{det_des}

To minimise electronic noise and suppress the signal from the laser repetition rate while still providing a high measurement bandwidth, we custom-build a photodetector designed for resonant detection at frequencies around the modulation sideband frequency. 
Resonant detection is achieved using resonant feedback in a transimpedance amplifier design, as shown in the circuit diagram in Fig.~\ref{fig:detector}.
This has several advantages over more commonly used designs, as discussed below.

To avoid saturation due to the large photocurrents, non-resonant detection of pulsed light generally requires a balanced photodiode configuration which subtracts the photocurrents produced by two photodiodes. This cancels both the mean photocurrent and the transient photocurrent at the laser repetition rate,
that would otherwise be present. Using resonant feedback eliminates the need for this subtraction by suppressing the response of the photodiode both at DC and above the resonance frequency. This makes it well suited for high power applications that use a single photodiode. Furthermore, compared with typical non-resonant designs based on $\sim 50~\Omega$ resistive loads, the $\sim 3$~k$\Omega$ resonant feedback impedance converts the photocurrent generated by the detector into a larger voltage.  This reduces 
requirements on the analog filters and buffer amplifiers needed to maintain signal integrity. 

Alternative resonant 
designs use the capacitance of the photodiode paired with a serial inductance to achieve both low noise and high sensitivity on resonance. However, since high intensity ps-pulses create free charges on the diode faster than they can be transported, the capacitance increases as the voltage across the  photodiode drops. The increased capacitance after the pulse leads to a significant shift of the resonance frequency. As this shift depends on the intensity of the pulse, such a design would require adjustments of either the signal frequency or the bias voltage to maintain the resonance condition at different intensity levels.

\begin{figure}[]
\begin{center}
\includegraphics[width=0.9\textwidth]{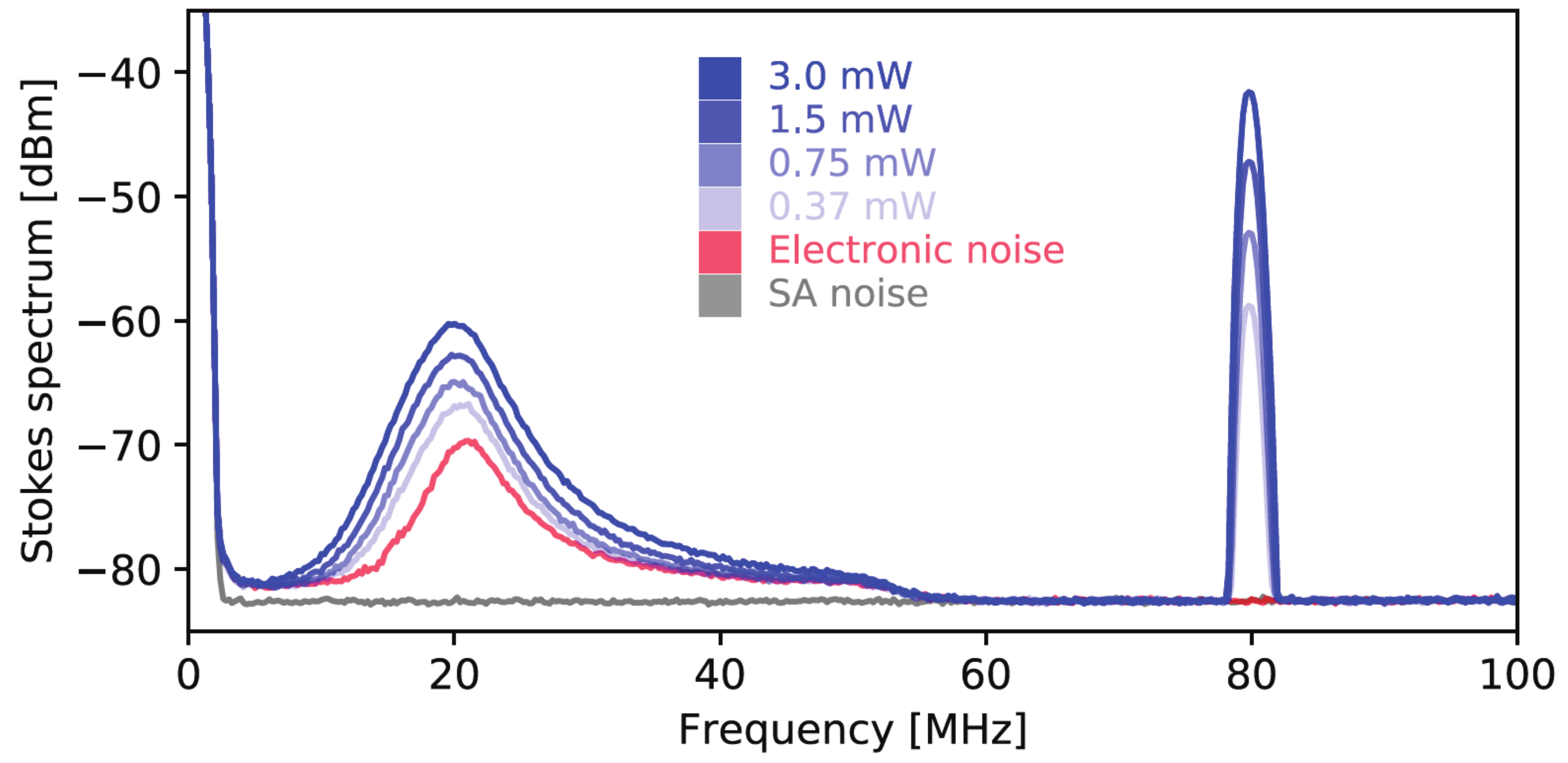}
\end{center}
\caption{\label{fig:det_noise}  Raw power spectral density of photocurrent from custom built resonant detector with no light on the detector (orange), and with increasing Stokes power levels (blue). Grey: spectrum analyser noise floor. RBW: 1 MHz, VBW: 300 Hz.}
\end{figure}

Since detection efficiency has a direct influence on the achievable quantum enhancement, a InGaAs-PIN photodiode was chosen for the detector. This choice of material generally offers a higher efficiency at a wavelength of 1064~nm than alternative IR-enhanced Si-PIN photodiodes. 
The higher capacitance per area and lower maximum reverse voltage apply stronger restrictions to the electronic circuit compared to Si photodiodes. However,  these photodiode parameters are less critical for resonant feedback detector than other resonant designs since the resonance frequency is determined predominantly by the feedback.  State-of-the-art stimulated Raman microscopes require photodetectors that have high power handling to maximise the useable Stokes laser power and therefore signal-to-noise, high bandwidth to maximise the achievable imaging frame rate, and low electronic noise~\cite{saar2010video}.  As a compromise between these requirements, we choose a  photodiode of 250~$\mu$m diameter, with larger diameter allowing better power handling at the cost of slower response and higher electronic noise.  

\subsection{Detector performance}

The raw power spectra observed for different Stokes power levels in the absence of squeezing are compared to the electronic noise floor of the detector in Fig.~\ref{fig:det_noise}. When detecting 3~mW of Stokes light we find that laser noise dominates electronic noise over a wide frequency band, from 9 to 40 MHz. The peaks in the electronic noise and Stokes noise variances at around 21 MHz are due to the resonant design of the photodetector. The sharp peak in the Stokes variance at 80 MHz corresponds to the repetition rate of the laser.

\subsection{Characterisation of quantum-enhancement}
\label{charact}

\begin{figure}
\begin{center}
\includegraphics[width=0.85\textwidth,keepaspectratio]{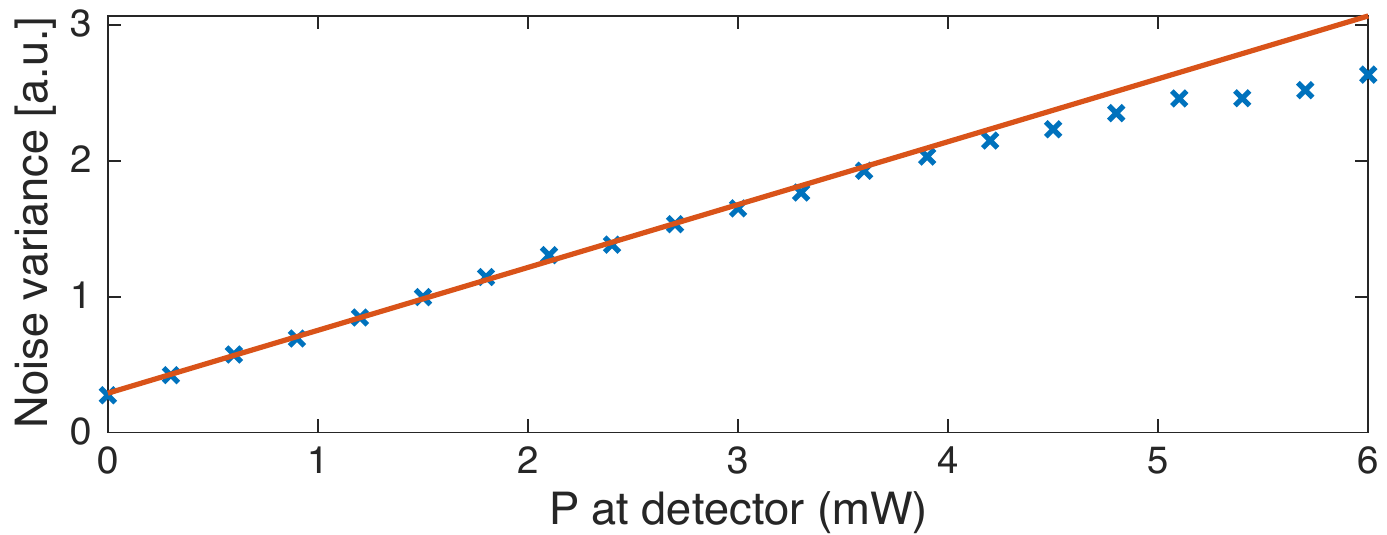} 
\caption{\textbf{Stokes laser is shot-noise limited.}  
 Power scaling of Stokes laser noise variance without squeezing, detected with our custom-built resonant feedback stimulated Raman photodetector. The variance is measured at the stimulated Raman modulation sideband frequency of 20.0083~MHz using a spectrum analyser with 1~kHz resolution bandwidth, matching the resolution bandwidth used to take data in the main text. Red line: linear fit to the expected scaling for shot-noise, excluding measurement points at high powers where the photodetector saturates. 
}
\label{figshot}
\end{center}
\end{figure}

Throughout the paper, the noise reduction due to quantum correlations was assessed by analysing the noise variance of the photocurrent from the resonant feedback photodetector using a spectrum analyser. The noise was calibrated against shot-noise using the Stokes laser with squeezer switched off as a reference, since this was known to be shot-noise limited (see below). To ensure a fair comparison, the same Stokes power of 3~mW is used at the photodetector with both squeezer on and off. All reported noise floor reductions are raw: that is, the electronic noise of the photodetector and spectrum analyser is not removed. As such, the measurements represent the actual useful improvement in noise floor that is achieved, rather than the ideal improvement that would be possible if there was no electronic noise.

The most direct method to determine the quantum enhancement of the stimulated Raman microscope would be to compare identical measurements with and without squeezing of the Stokes field. The Raman scattering signal depends strongly on the spatiotemporal modeshape of the Stokes field, since this affects both the Stokes intensity at the sample and the spatiotemporal overlap of the Stokes field with the Raman pump. Consequently, if the comparison involved the Raman signal (for instance, a comparison of signal-to-noise ratios), it would be  crucial to ensure that the spatiotemporal modeshape of the Stokes field remain unchanged when the squeezing is switched on -- any modification of the spatiotemporal modeshape would alter the Raman signal size in a purely classical way, which should not be misinterpreted as a quantum enhancement (or dehancement) due to the presence of correlated photons. Maintaining identical spatiotemporal profiles with and without squeezing, proved not to be possible in practise due to the non-perfect spatiotemporal overlap between the 532~nm OPA pump laser and the Stokes field within the squeezing crystal. The effect of this non-perfect overlap is explored in some detail in Section~\ref{model} below. The important concept, here, is that only the component of the Stokes field that is overlapped with the OPA pump is deamplified, with the non-overlapped component left to transmit through the crystal unaffected. When the squeezer is switched on, the deamplification reduces the overlapped Stokes intensity by more than 50\% (see Fig.~\ref{fig:sqz_perf}{\bf b}) but leaves the non-overlapped component unchanged. This grossly changes the spatiotemporal modeshape of the Stokes field.

To achieve a fair comparison, we instead only measure the Raman signal size $\rm S$ with the squeezer on. We assume that, for identical Stokes spatiotemporal modeshape, the signal when using shot-noise limited light will have the same magnitude. This is an appropriate assumption, since the scattering cross-section, and therefore number of pump photons Raman scattered into the Stokes field, is independent of quantum correlations between photons in the Stokes field. We note that this assumption, and approach, has been used to characterise the quantum enhancement in other work for similar reasons~\cite{triginer2020quantum}. The quantum enhanced signal-to-noise ratio in Fig. 3{\bf c} of the main text is then be directly calculated as $\rm SNR^{quantum} = S/N^{\rm quantum}$, where $\rm N^{\rm quantum}$ is the noise floor of the quantum-enhanced measurement; with the shot-noise limited signal-to-noise calculated as $\rm SNR^{shot-noise} = S/N^{\rm shot-noise} = SNR^{\rm quantum} \times N^{\rm quantum}/N^{\rm shot-noise}$, where $\rm N^{\rm quantum}$ is the noise floor of the shot-noise limited measurement. Similarly, for the imaging in Fig.~4 of the main text, the image is normalised to the shot-noise floor by dividing the Raman signal observed with quantum-enhancement by the shot-noise floor determined by both switching the squeezer off and blocking the Raman pump field.

\subsection{Verification that the Stokes laser is shot-noise limited}
\label{verif}

To verify that  the Stokes laser is shot-noise limited when the squeezer switched off, as is critical when using it for noise calibration,
we measure the variance of its noise as a function of power. This is shown in the power spectra in Fig.~\ref{fig:det_noise}. Over the full range of frequencies where the light noise dominates the electronic noise, the light noise was found to scale linearly with Stokes power for powers beneath detector saturation. This linear scaling is consistent with expectations for a shot-noise limited laser. If the laser were instead limited by technical noise, such as laser relaxation oscillation or spontaneous emission noise, quadratic scaling would be expected. 

Most important for the experiments reported here, is that prior to squeezing the Stokes light is shot-noise limited around the stimulated Raman modulation frequency. Measurements at that frequency are shown as a function of power 
in Fig.~\ref{figshot}. Linear scaling is observed for optical powers below 4 mW, with detector saturation occurring at higher powers. 
As can be seen, that data is inconsistent with the quadratic scaling expected for technical noise.
 As such it may be concluded that with the squeezer switched off the Stokes laser is shot-noise limited at the 3~mW power levels used in our experiments.

\section{Measurement and validation of squeezing}

\subsection{Observed squeezing spectrum}

\begin{figure}[]
\begin{center}
\includegraphics[width=0.9\textwidth]{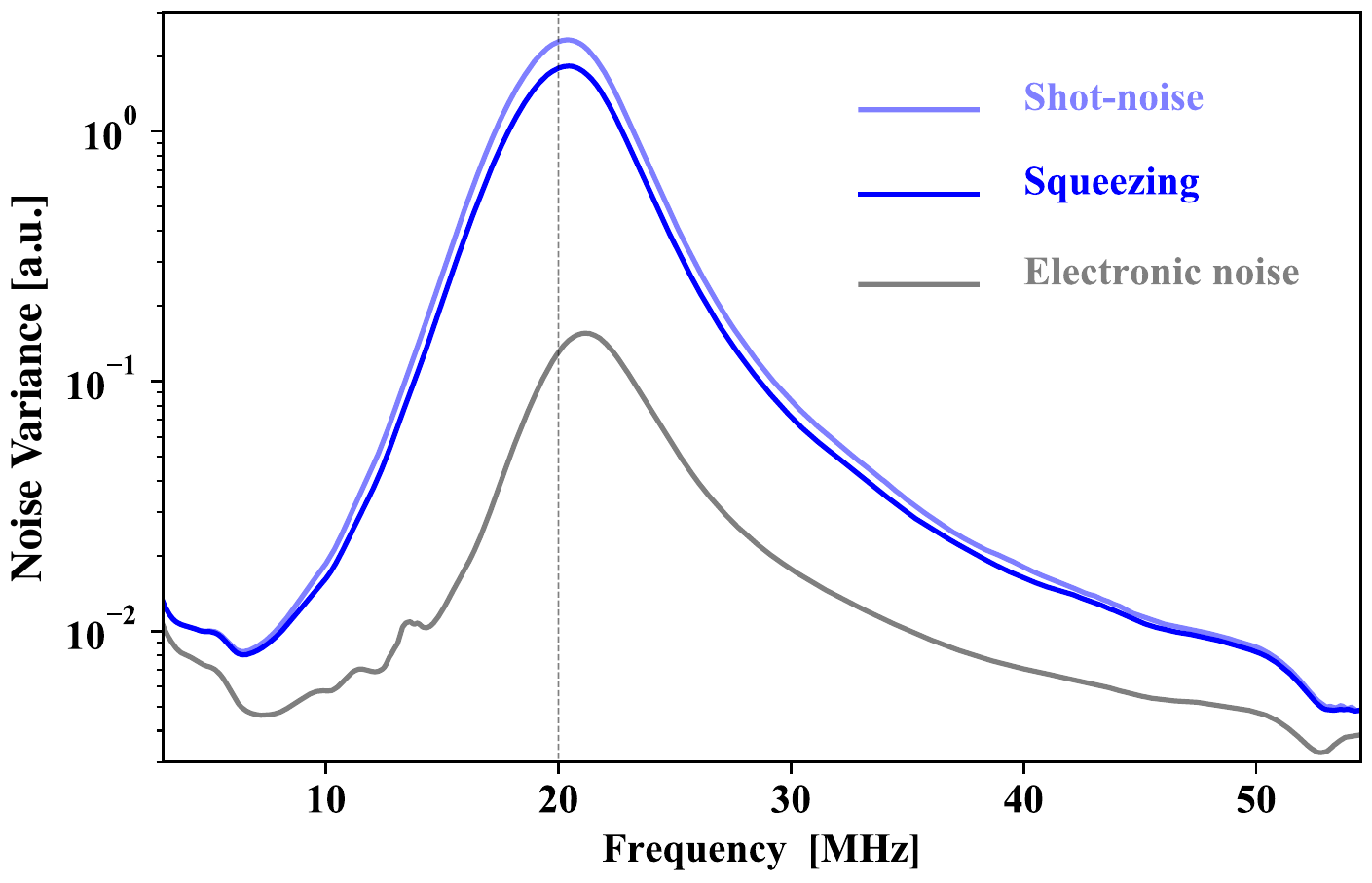}
\end{center}
\caption{\label{fig:det_freq}  Raw power spectral density of photocurrent from custom built resonant detector with no light on the detector (grey), 3 mW of shot-noise limited light on the detector (light blue), and 3 mW of amplitude squeezed light on the detector (dark blue). RBW: 1 MHz. VBW: 10 Hz. Averaged thirty five times.
}
\end{figure}

Fig.~\ref{fig:det_freq} shows a typical observed raw power spectrum of the squeezed light generation in our OPA, compared to both the electronic noise floor of the detector and the shot-noise of the Stokes laser observed with the same 3 mW of power incident on the detector. It can be seen that the noise on the squeezed field is suppressed compared to the shot-noise limit over a wide frequency range. Fig,~3{\bf a,} in the main text normalises the squeezing to the shot-noise level to show the quantum noise reduction more clearly as a function of frequency. In this figure, and for all reported quantum-enhancement values, electronic noise is not corrected for; that is, the quantum enhancement reported is the real enhancement of the measurement, not the enhancement that would be achieved if the detector was replaced with an ideal detector with no electronic noise. As can be seen in that figure, suppression of the noise beneath the shot-noise level is achieved for frequencies between 5 MHz and 52 MHz, though it is degraded at the extremes due to the electronic noise of the detector.  As such, the quantum enhancement we observe could be applied in scenarios with sub-microsecond pixel dwell times, allowing video rate stimulated Raman imaging. 

\subsection{Modelling the squeezing process}
\label{model}

\begin{figure}[b!]
\begin{center}
\includegraphics[width=0.9\textwidth]{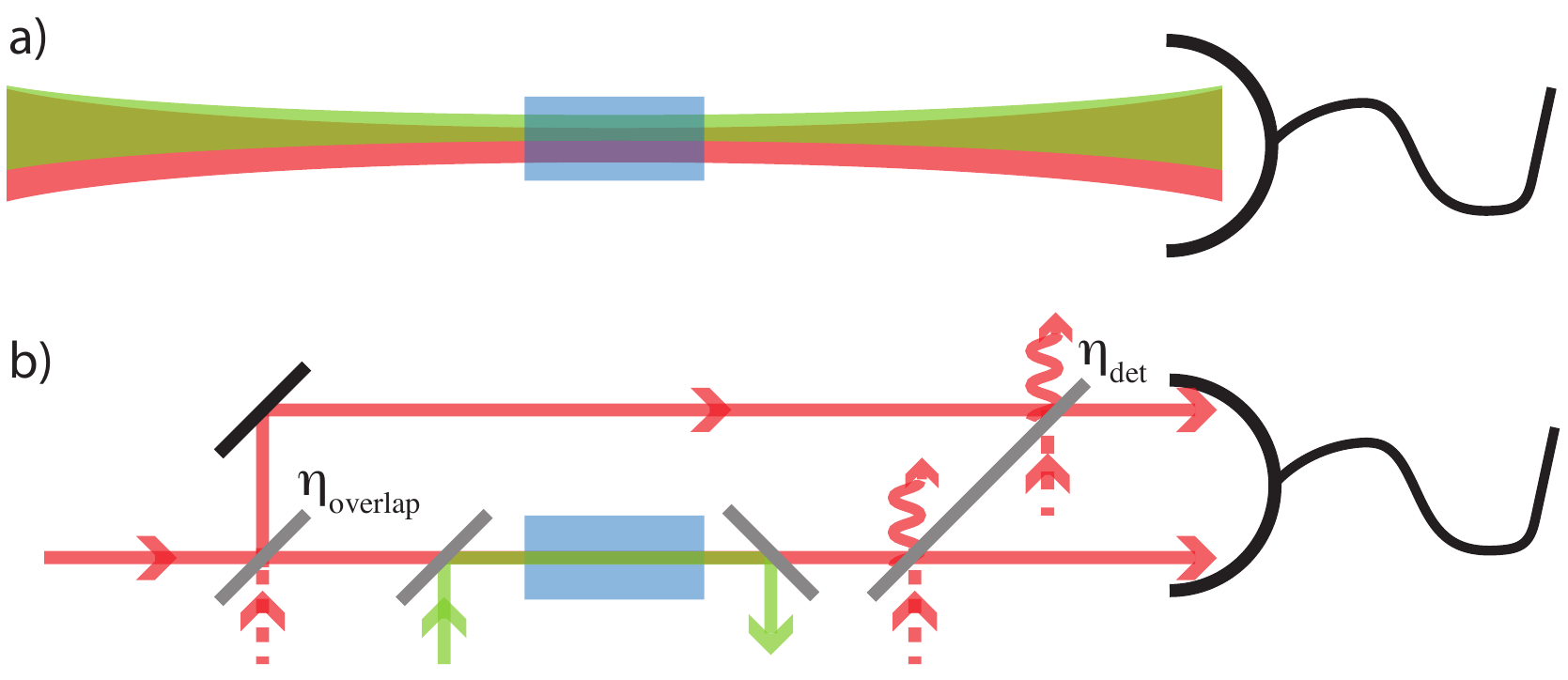}
\end{center}
\caption{\label{fig:theory} 
}
\end{figure}

As described above and in the main text, to produce squeezed light we utilise a single pass optical parametric amplifier. The simplicity of a single-pass configuration makes it technically desirable from alignment, construction and control perspectives. It is possible here, since the use of picosecond pulses increases the peak intensity of the light, and therefore the nonlinear interaction, making resonant enhancement in a cavity redundant. However, the lack of an optical cavity also introduces the possibility of multi-spatial mode effects which could potentially deleteriously effect the level of squeezing generated. To investigate this, and confirm that the squeezing generation is well understood, we develop consider a simple model of the squeezing process, taking account of non-perfect spatial overlap between the OPA pump field and the Stokes field. 

The concept of the model is shown in Fig.~\ref{fig:theory}. Fig.~\ref{fig:theory}a shows a rough schematic of the squeezer with OPA pump and Stokes partially misaligned. Only the spatially overlapped component of the Stoke laser will experience amplification due to the pump. The non-overlapped component will pass through the nonlinear crystal unaffected. Fig.~\ref{fig:theory}b shows a basic model of this, with a fraction $\eta_{\rm overlap}$ of the Stokes field perfectly overlapped with the OPA pump, and a fraction $1-\eta_{\rm overlap}$ bypassing the pump entirely. Both component of the field them propagate towards the photodetector, and reach it with the same transmission efficiency $\eta_{\rm det}$ (including the inefficiency of the photodetector itself). They are then detected together on the photodetector.

The amplitude $\hat X_1$ and phase $\hat Y_1$ quadratures of the component of the detected light that does not experience squeezing are given by
\begin{eqnarray}
\hat X_1 &=& \sqrt{\eta_{\rm det}} \left ( \sqrt{1-\eta_{\rm overlap}} \hat X_{\rm in} - \sqrt{\eta_{\rm overlap}} \hat X_{v1} \right ) +  \sqrt{1- \eta_{\rm det}} \hat X_{v2} \label{q1}\\
\hat Y_1 &=& \sqrt{\eta_{\rm det}} \left (  \sqrt{1-\eta_{\rm overlap}} \hat Y_{\rm in} - \sqrt{\eta_{\rm overlap}} \hat Y_{v1} \right ) +  \sqrt{1- \eta_{\rm det}} \hat Y_{v2},
\end{eqnarray}
where $\hat X_{\rm in}$ and $\hat Y_{\rm in}$ are the amplitude and phase quadratures of the input Stokes light, and the subscripts $v1$ and $v2$  label the vacuum noise entering due to poor overlap and transmission inefficiency, respectively.

Similarly, the amplitude $\hat X_2$ and phase $\hat Y_2$ quadratures of the component of the detected light that is perfectly overlapped to experience squeezing are given by
\begin{eqnarray}
\hat X_2 &=& \sqrt{G \eta_{\rm det}} \left ( \sqrt{\eta_{\rm overlap}} \hat X_{\rm in} + \sqrt{1-\eta_{\rm overlap}} \hat X_{v1} \right ) +  \sqrt{1- \eta_{\rm det}} \hat X_{v3} \label{q2} \\
\hat Y_2 &=& \sqrt{\frac{\eta_{\rm det}}{G}} \left (  \sqrt{\eta_{\rm overlap}} \hat Y_{\rm in} + \sqrt{1-\eta_{\rm overlap}} \hat Y_{v1} \right ) +  \sqrt{1- \eta_{\rm det}} \hat Y_{v3},
\end{eqnarray}
where the subscript $v3$ labels the vacuum noise entering due to  transmission inefficiency for this field, and $G$ is the level of squeezing of the variance of the perfectly overlapped squeezed light, prior to detection inefficiencies. Here, $G<1$ for amplitude squeezing.

Since the perfectly overlapped and non-overlapped fields are spatially orthogonal, when they are detected together on the photodiode they do not interfere, so that the resulting photocurrent is simply the sum of the photocurrents that would be measured if each was detected independently:
\begin{eqnarray}
i &=& a_1^\dagger a_1 + a_2^\dagger a_2\\
&=& \alpha_1^2 + \alpha_2^2 + \alpha_1 \delta \hat X_1 + \alpha_2 \delta \hat X_2, \label{asf}
\end{eqnarray}
where the $a$'s are annihilation operators for the fields, with $\hat X = a + a^\dagger$ and $\hat Y = i (a^\dagger - a)$, the coherent amplitude is defined as $\alpha = \langle a \rangle$ and is defined here to be real without loss of generality, and the noise operators are defined as $\delta \hat X = \hat X - \langle \hat X \rangle$ so that $\langle \delta \hat X \rangle =0$. Here, to get to Eq.~(\ref{asf}) we have used the usual linearisation approximation, assuming that $\{ \alpha_1, \alpha_2 \} \gg 1$ so that noise operator product terms can be neglected.

The squeezer acts to both deamplify the coherent amplitude of the Stokes field and decrease its noise variance. The deamplification is given by
\begin{equation}
D = \frac{\langle i \rangle}{\langle i \rangle_{G=1}} = \frac{ \alpha_1^2 + \alpha_2^2}{ \alpha_1^2 + \alpha_{2,G=1}^2}. \label{d}
\end{equation}
The fractional reduction in noise variance beneath the shot-noise level is 
\begin{equation}
V = \frac{\langle i^2 \rangle - \langle i \rangle^2}{(\langle i^2 \rangle - \langle i \rangle^2)_{G=1}} = \frac{\alpha^2_1 + \alpha_2^2 V_2}{\alpha^2_1 +\alpha^2_2 } \label{v}
\end{equation}
where we have used the fact that the noise operators of the perfectly-overlapped and non-overlapped fields are uncorrelated, have defined the shot-noise limited variance of the non-overlapped field $V_1 = \langle \delta \hat X_2^2 \rangle$ to equal one, and $V_2 = \langle \delta \hat X_2^2 \rangle$ is the variance of the detected perfectly-overlapped squeezed component in the absence of the non-overlapped field. We have also used the fact that when the squeezer is switched off ($G=1$) the overlapped field is shot-noise limited so that $V_2=1$.

To determine the effect of mode mismatch and detection inefficiency on the observed amplification and squeezing, we therefore need to determine $\alpha_1$, $\alpha_2$ and $V_2$. From Eqs.~(\ref{q1})~and~(\ref{q2}), the detected coherent amplitudes of the non-overlapped and perfectly overlapped fields when detected can be shown to be
\begin{eqnarray}
\alpha_1 &=& \sqrt{\eta_{\rm det} (1-\eta_{\rm overlap})} \alpha_{\rm in}\\
\alpha_2 &=& \sqrt{G \eta_{\rm det} \eta_{\rm overlap}} \alpha_{\rm in},
\end{eqnarray}
and the squeezing of the perfectly overlapped field to be
\begin{equation}
V_2 = G \eta_{\rm det} + 1- \eta_{\rm det},
\end{equation}
where we have appropriately taken the input field and vacuum noise inputs to be shot-noise limited ($\langle \delta \hat X_{\rm in}^2 \rangle =\langle \delta \hat X_{v1}^2 \rangle  = \langle \delta \hat X_{v3}^2 \rangle = 1$) and uncorrelated with each other. 

Substituting these expressions into Eqs.~(\ref{d})~and~(\ref{v}) we find
\begin{equation}
D = 1 - \eta_{\rm overlap} + G \eta_{\rm overlap}, \label{dd}
\end{equation}
and
\begin{equation}
V = \eta_{\rm det} \left (\frac{1 - \eta_{\rm overlap} + \eta_{\rm overlap} G^2 }{1 - \eta_{\rm overlap} + \eta_{\rm overlap} G} \right ) + 1 - \eta_{\rm det}. \label{vv}
\end{equation}
As expected, we see that the deamplification is independent of detection efficiency - the detection inefficiency affects the coherent amplitude in exactly the same way independent of whether the light is squeezed or not. As $G \rightarrow 0$ the deamplification asymptotes to $1-\eta_{\rm overlap}$. This makes sense, since in that limit the spatially overlapped fraction of the light is perfectly deamplified, while the non-overlapped component remains unaffected.

 The predicted noise variance $V$ exhibits behaviour unexpected from a normal single-mode squeezer. Instead of the variance decreasing monotonically as $G \rightarrow 0$ from above, it decreases below the shot-noise limit  and then rises again to exactly the shot-noise limit ($V=1$) at $G=0$. This can be understood since as discussed above, in this limit the coherent amplitude of the squeezed component goes to zero. Since the magnitude of the detected photocurrent scales with the coherent amplitude, the squeezed-light signal in the photocurrent is fully suppressed, only leaving the signal from the shot-noise limited field that has no overlap.
 
 It is straightforward to calculate the gain at which the squeezing is maximised as well as the maximum squeezing. Doing this, we find
 \begin{eqnarray}
 G_{\rm min \,V} &=& \frac{\sqrt{1-\eta_{\rm overlap}}}{\eta_{\rm overlap}} \left (1 - \sqrt{1-\eta_{\rm overlap}} \right )\\
 V_{\rm min} &=& 2G_{\rm min \,V} = \frac{2 \sqrt{1-\eta_{\rm overlap}}}{\eta_{\rm overlap}} \left (1 - \sqrt{1-\eta_{\rm overlap}} \right ).
 \end{eqnarray}

Assuming no pump depletion, which is reasonable in our case, the seed light (the input Stokes field) is exponentially amplified as it propagates through the OPA crystal to produce the gain $G$ on the final output amplitude quadrature (and 1/$G$ on the phase quadrature). The gain can then be expressed as $G= e^{-\kappa L}$ where $\kappa$ is the gain per unit length and $L$ is the length of the crystal. Recognising that the nonlinear interaction is proportional to the amplitude of the OPA pump field, so that $\kappa \propto \sqrt{P}$ where $P$ is the OPA pump power, this can be re-expressed as 
 $G = e^{-g \sqrt{P}}$, where 
 $g$ is a gain coefficient that depends on the length, phase matching and nonlinear coefficient of the crystal, as well as the optical focussing within the crystal. With this relationship it is possible to compare the predictions of the model to
 experiment. 
 
 \subsection{Comparison of model and experiment}
 
 \begin{figure}[]
\begin{center}
\includegraphics[width=0.85\textwidth]{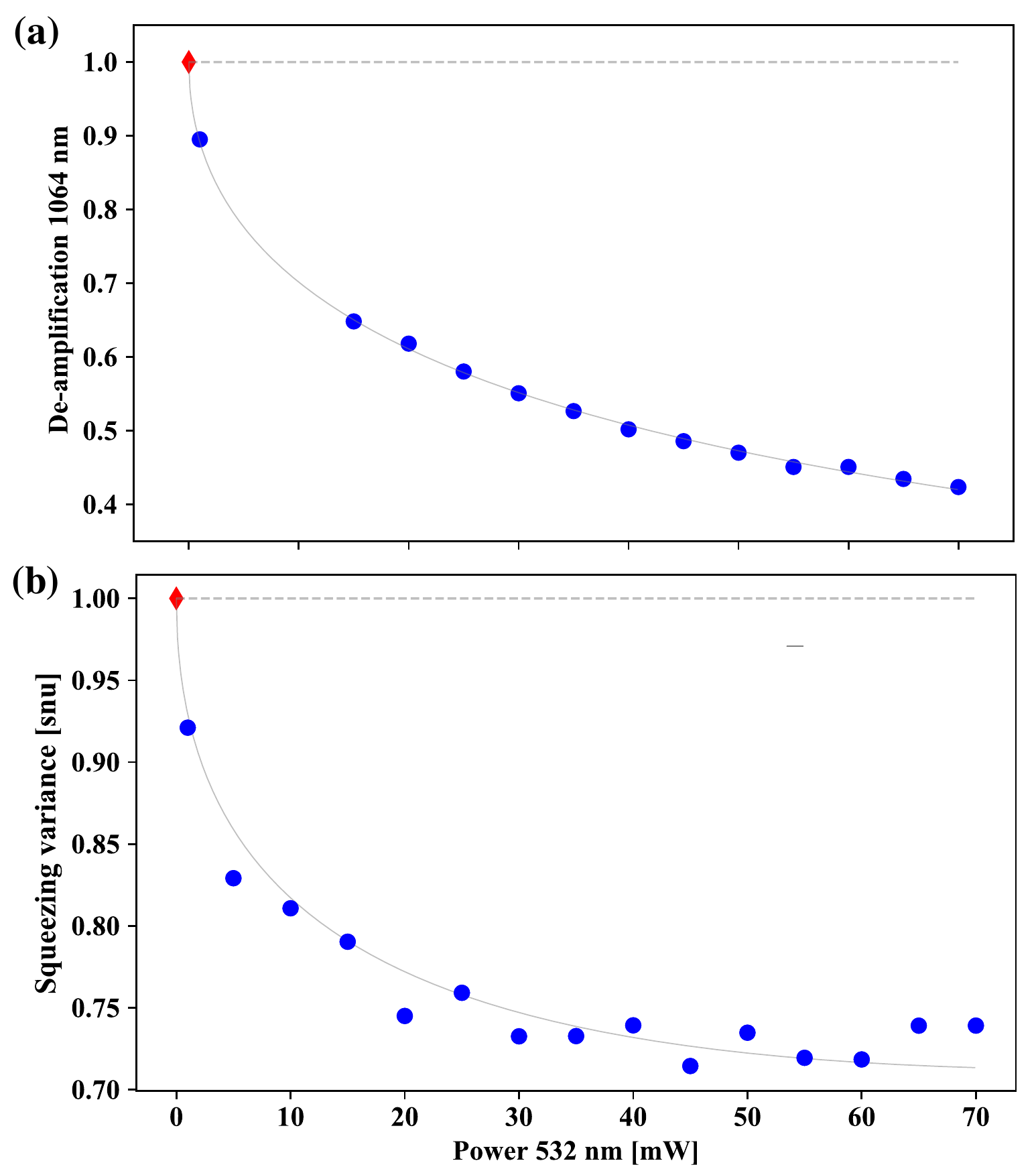}
\end{center}
\caption{\label{fig:sqz_perf}  Validating the model and characterizing the performance of the squeezer. (a) Measured reduction in the intensity of the seed field after the OPA as a function of OPA pump power. Line: least squares fit to data with the overlap efficiency $\eta_{\rm overlap}$ and squeezer gain $g$ as fitting parameters. The fit yielded values of $\eta_{\rm overlap} = 0.85 \pm 0.02$ and $g = 8.8 \pm 0.4$~W$^{-1/2}$. (b)  Observed squeezed variance at 20~MHz as a function of OPA pump power. Line: least squares fit to data with only the detection efficiency $\eta_{\rm det}$ as a free parameter,  $\eta_{\rm overlap} = 0.85$, and $g = 8.8$~W$^{-1/2}$. The fit yielded $\eta_{\rm det} =0.65 \pm 0.01$. snu: shot-noise units.
}
\end{figure}
 
 To compare model and experiment we measure both the deamplification and squeezing as the pump power is varied from 0 to 70~mW. The results are shown in Fig.~\ref{fig:sqz_perf}. We fit the deamplification to Eq.~(\ref{dd}) extracting the overlap efficiency $\eta_o = 0.85 \pm 0.02$ and gain coefficient $g = 8.8 \pm 0.4$~W$^{-1/2}$. Fixing these parameters and fitting the squeezed variance to Eq.~(\ref{vv}) we determine the detection efficiency in this case (without passing through the microscope) to be $\eta_d=0.65 \pm 0.01$, consistent with expectations given the measured 72\% efficiency of the detector and  inefficiencies in transmission from the nonlinear crystal to the detector. As can be seen from Fig.~\ref{fig:sqz_perf}, the data agrees well with the theory\footnote{We note that the variance data disagrees with more basic theory that does not account for the effect of deamplification of the coherent amplitude of the overlapped field on the detected variance.}, validating this simple model. The agreement also provides further validations that, prior to the squeezing process, the Stokes field is shot-noise limited and that our method of determining the quantum enhancement is accurate.

As can be seen in Fig.~\ref{fig:sqz_perf} the variance of the detecter squeezing is optimised and relatively insensitive to OPA pump power in the power range between 35 and 65 mW. We therefore operate our experiments in this regime. 

\section{Total detection efficiency of squeezed light}

Optical inefficiencies reduce quantum correlations between photons and therefore degrade the level of quantum enhancement that is possible using squeezed light. The path between the nonlinear crystal generating squeezing and eventual photodetection, included a range of optical elements which introduce loss, as shown schematically in Fig. 1{\bf a} of the main text. The microscope itself was a significant source of loss, in the objectives, in dichroic mirrors, and due to reflections from the microscope coverslips. The total efficiency of transmission through the microscope was measured to be $\eta_{\rm microscope} = 0.75$.  The mirrors, lenses and dichroic mirrors between the nonlinear crystal and microscope and between the microscope and photodetector were measured to have transmission of $\eta_{\rm propagation} = 0.82$. The photodetector efficiency was determined by measuring photocurrent $i$ produced by $P_{\rm Stokes} = 3$~mW of Stokes light and comparing the flux of electrons this corresponds to to the flux of photons arriving at the detector. The measured photocurrent of $i = 1.86$~mA resulted in a detector quantum efficiency of
\begin{equation}
\eta_{\rm detector} = \frac{\hbar \omega}{P_{\rm Stokes}} \times \frac{i}{e} = 0.72,
\end{equation}
where $e$ is the charge on an electron, $\hbar$ is the reduced Plancks constant, and $\omega = 2 \pi c/\lambda$ is the angular optical frequency, with $c$ the speed of light and $\lambda=1064$~nm the Stokes wavelength. Combined, this gives a total efficiency of transmission of the squeezed light from the nonlinear crystal to the measured photocurrent of
\begin{equation}
\eta = \eta_{\rm microscope} \eta_{\rm propagation} \eta_{\rm detector}  = 0.44.
\end{equation}

It is interesting to ask whether the measured squeezing after passing the squeezed field through the microscope is consistent with this measured total efficiency. The theoretical squeezed variance after transmission with an efficiency of $\eta$ is $V_{\rm sqz}^{\rm after} = \eta V_{\rm sqz} + 1 - \eta$, where $V_{\rm sqz}$ is the squeezed variance directly after the OPA crystal. $V_{\rm sqz}$ can be estimated from the measured squeezed variance prior to the microscope $V_{\rm sqz}^{\rm prior}$ using then relation $V_{\rm sqz}^{\rm prior} = \eta_{\rm det} V_{\rm sqz} + 1 - \eta_{\rm det}$, with the efficiency $\eta_{\rm det} = 0.65$ determined in Section~\ref{model}. Under ideal operating conditions, $V_{\rm sqz}^{\rm prior} = 0.72$ (or -1.4 dB), as shown in Fig.~\ref{fig:sqz_perf}. However,  the operating conditions of the squeezed light source vary from experiment to experiment due to variations in the overlap between the OPA pump and the input Stokes field, and variation in the performance of the phase lock. This results in degraded performance in some experiments (for instance, see Fig.~3{\bf a} in the main text, for which $V_{\rm sqz}^{\rm prior} = 1 - 0.22 = 0.78$). Overall, we find that from experiment to experiment $V_{\rm sqz}^{\rm prior}$ varies in the range $V_{\rm sqz}^{\rm prior} = 0.76 \pm 0.04$. From this, we find the range of variances for the amplitude squeezing directly after the OPA crystal of $V_{\rm sqz} = 0.63  \pm 0.06$. After transmission through the microscope apparatus and photodetection with a combined efficiency of $\eta$, the predicted detected squeezing is $V_{\rm sqz}^{\rm after} = 0.84 \pm 0.03$. This is broadly consistent with our measurements after the microscope, with squeezed variance in Fig.~3{\bf b} of the main text equal to 0.87, that in the quantum image of polystyrene beads in Fig.~4{ \bf a} of the main text equal to 0.81. In the case of live cell imaging, significant efforts were made to improve the alignment of the system, including improvements in the overlap between pump and seed in the optical parametric oscillator generating squeezing, of the alignment of pump and Stokes in the Raman microscope, and of the microscope itself. Fresnel reflective losses were also lower at the water-glass  interfaces between sample and coverslips, than at the air-glass interfaces when imaging dry polystyrene samples. This resulted in an improved squeezed variance of 0.74 (Fig.~4{\bf b}).

\section{Estimation of lower limit to concentration sensitivity}

The lower limit to detectable concentration is reached where SNR=1. Since the SNR increases with averaging time, we use normalized units, equivalent to the sensitivity achievable for the case of 1~s acquisition time. It is possible to estimate this concentration by extrapolating how the measured SNR changes as the acquisition time changes and the concentration changes. The stimulated Raman signal amplitude scales linearly with concentration $C$. Since the SNR is linearly proportional to the power of the stimulated Raman signal, it therefore scales quadratically with concentration .Hence, the minimum detectable concentration is $C_{\rm min} = C/$SNR$^{1/2}$. The SNR increases linearly with time so that the SNR in normalized units can be related to SNR($\tau$) for an arbitrary  acquisition time $\tau$ via SNR(1~s)=SNR$(\tau)\times$(1~s)$/\tau$. We therefore find that the minimum detectable concentration can be found as $C_{\rm min} = C/$SNR$(\tau)^{1/2}\times [\tau/$(1~s)$]^{1/2}$.

Polystyrene is a polymer formed from the monomer styrene, and does not have a well-defined molecular weight or number of aromatic rings. However every aromatic ring corresponds to one styrene monomer. Using the density of polystyrene, 1060~kg/m$^3$, and the molecular weight of styrene, 104.15~g/mol, we therefore find the polystyrene concentration to be $C=10.2$~M. Note that the unit M corresponds to mol/L. 

The damage free shot-noise limited SNR was 88 (Fig.~3{\bf c} of the main text), recorded with 3~kHz resolution bandwidth which corresponds to an effective measurement time of $\tau=1/3000$~s. Extrapolating to SNR=1 with 1~Hz resolution bandwidth, the minimum concentration sensitivity is estimated as $C_{\rm min}=19.8$~mM Hz$^{-1/2}$. The quantum enhanced SNR was 99, also recorded with 3~kHz resolution bandwidth. Similarly extrapolating this to SNR=1 provides an estimated minimum concentration sensitivity of $C_{\rm min}=18.7$~mM Hz$^{-1/2}$. To allow comparison with other stimulated Raman microscopes, we also consider that the minimum detectable concentration also scales linearly with pump power and as the square root of Stokes power, which in this case were 23~mW and 3~mW respectively. As such our minimum resolvable concentration can be normalized to $C_{\rm min}=745$~mM~Hz$^{-1/2}$~ mW$^{-3/2}$. 
 
\section{Imaging}

The quantum-enhanced images are formed by recording the stimulated Raman signal on the spectrum analyser while raster scanning the sample in 100~nm steps at a rate of 20 steps per second, returning a 100$\times$100 pixel image in 8.3 minutes. The stage position is recorded throughout the raster scan and is used to assign positions to the measured Raman values. After recording the stimulated Raman signal at each position, the shot-noise is also measured at each of the same locations by blocking the pump field, keeping the Stokes power at the detector at 3~mW, and recording shot-noise across a similar raster scan. The Raman signal is normalized by dividing by the shot-noise level at the same locations, which corrects for any change in signal strength due to attenuation of the detected Stokes field by the sample.

\subsection{Observation of yeast cell photodamage}

\begin{figure}
\begin{center}
\includegraphics[width=0.75\textwidth,keepaspectratio]{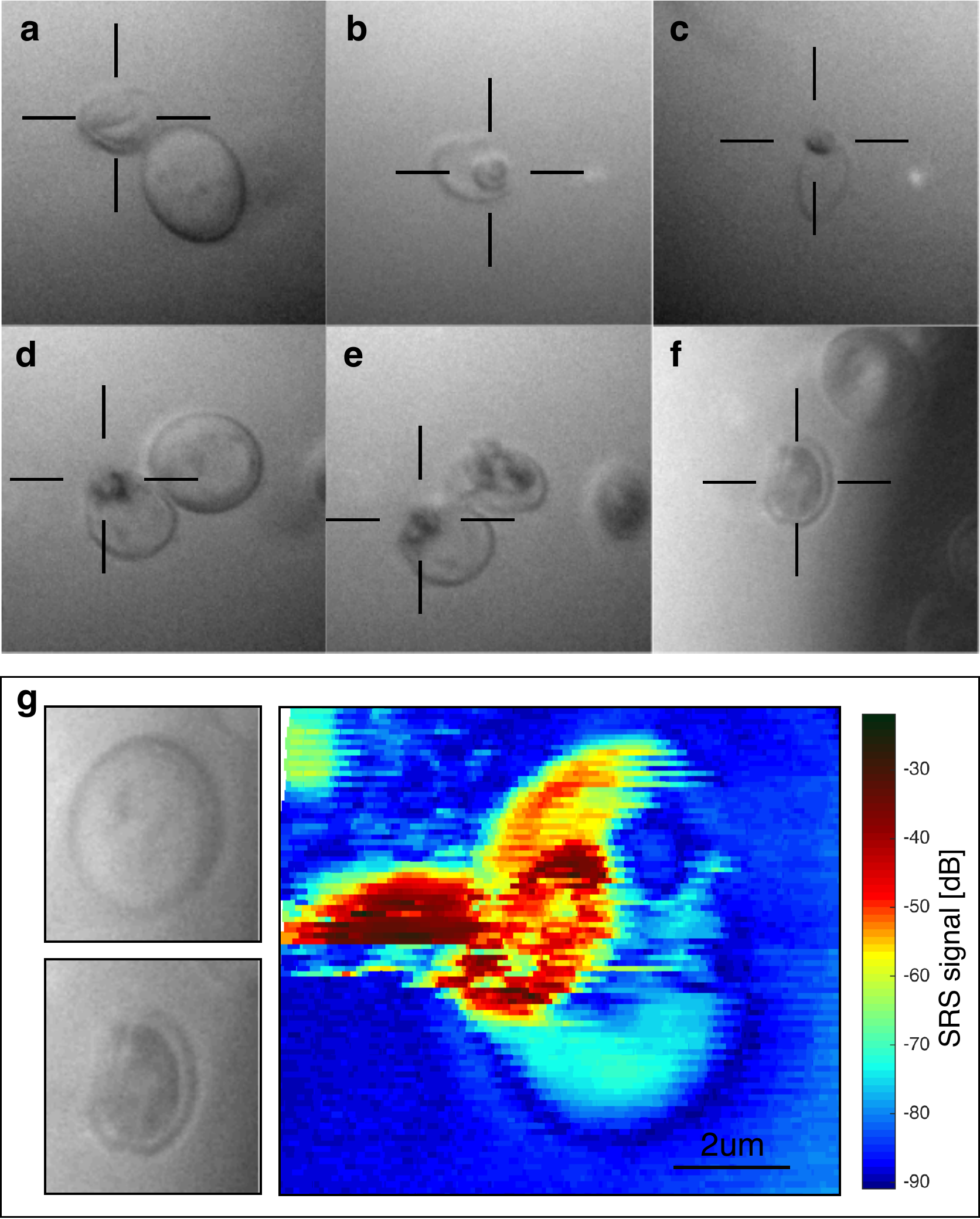} 
\caption{\textbf{Bright field images and Raman signal image of damaged yeast cells.}  
Photodamage induced by the pump field on yeast cells with 30~mW of pump power at the sample and an effective microscope objective numerical aperture of NA$_{\rm effective} \sim1.2$, corresponding to an intensity of around 470~W/$\mu$m$^2$. {\bf a} to  {\bf f}: widefield images of damaged cells. 
{\bf g}: Photodamage induced during simulated Raman imaging. Left: widefield images before (top) and after (bottom) damage. Right: Raman image showing damage occurring mid-scan.}
\label{cell_damage}
\end{center}
\end{figure}

In addition to quantifying the optical damage on polystyrene beads, we were able to observe and illustrate light-induced damage on the yeast cells. Photodamage appeared for intensities of 470~W/$\mu$m$^2$.
We document in Figure \ref{cell_damage}{\bf g} a case of photodamage occurring during a scan. The image was scanned in horizontal scans, stepping vertically between scans and starting an the bottom of the image. The signal amplitude increased considerably after the appearance of visible damage. The Raman signal increased by up to 40~dB and debris with high Raman signatures were dragged outside of the cell by the horizontal motion of the beam.
We also present in Figure~\ref{cell_damage} widefield images of damaged cells. The damage was often localized and associated with shrinking. This suggests that the light may have caused membrane destruction followed by diffusion of the organelles into the extracellular medium.

\subsection{Comparison of squeezed and shot noise limited images}

\begin{figure}
\begin{center}
\includegraphics[width=0.85\textwidth,keepaspectratio]{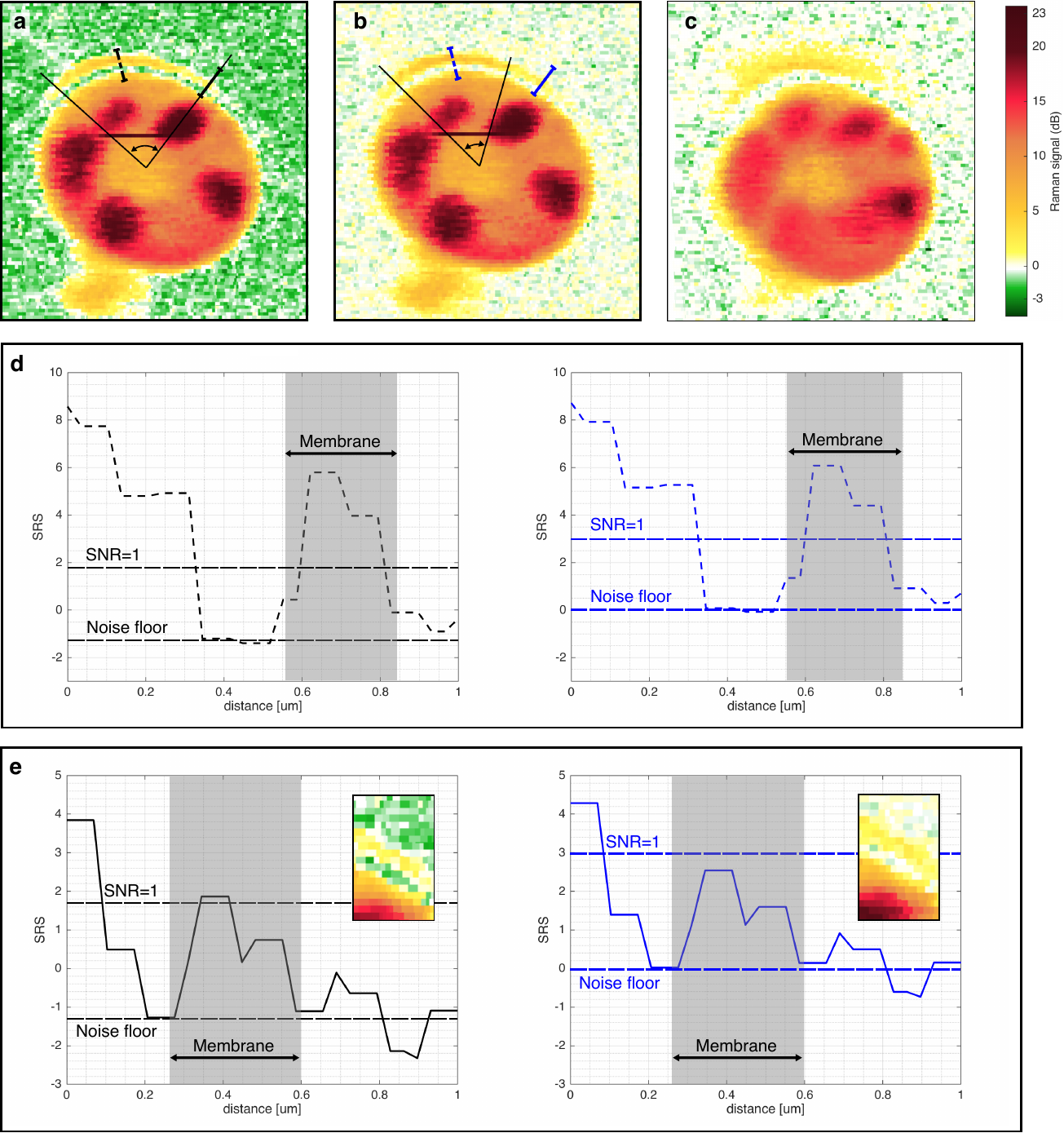} 
\caption{\textbf{Comparison of quantum-enhanced and shot-noise limited images.}
 {\bf a},{\bf b}~\&~{\bf c} respectively show the quantum-enhanced image, shot-noise limited image produced by artificially raising the background of the quantum enhanced image by 1.3~dB, and the shot-noise limited image. An interpolation of the Raman signal is performed radially from the center of the cell to quantify the resolvable fraction of the membrane.
 {\bf d} Raman signals along the dashed lines in  {\bf a}~\&~{\bf b}, corresponding to the cross-section with maximum membrane signal-to-noise. Here we see SRS signals are approximately 7 and 6~dB higher than the noise floor, respectively, for the quantum-enhanced and shot-noise limited images. {\bf e} Raman signals along the solid lines in {\bf a}~\&~{\bf b}, corresponding to cross-section with degraded membrane signal-to-noise. Here the SRS signals are are approximately 3.1 and 2.5~dB higher than the noise floors, respectively, for the quantum-enhanced and shot-noise limited images. In this case, only the quantum-enhanced measurements has a signal-to-noise above one. In  {\bf d}\&{\bf e}: insets are magnified parts of the Raman images, showing the membrane; grey shading shows the location of the membrane; upper dashed horizontal lines show a signal-to-noise of one; and lower dashed horizontal lines show the noise floor normalised to the shot-noise level.}
\label{image_comp}
\end{center}
\end{figure}

Figure~\ref{image_comp} compares a quantum enhanced cell image (Fig.~\ref{image_comp}{\bf a}), to the equivalent shot-noise limited image obtained by artificially raising the noise floor to the shot noise floor (Fig.~\ref{image_comp}{\bf b}), and to an experimental shot noise limited image (Fig.~\ref{image_comp}{\bf c}). The removal of quantum correlated light in Fig.~\ref{image_comp}{\bf b}\&{\bf c} results in an evident reduction in contrast. The experimental shot noise limited image was taken 30 minutes after the quantum-enhanced image. The sample was found to drift axially over these timescales. We attribute the degradation in resolution observable in Fig.~\ref{image_comp}{\bf c} to this drift, rather than the removal of quantum correlations.
 Fig.~\ref{image_comp}{\bf d} plots the Raman signal along the dashed cross sections on  Figs.~\ref{image_comp}{\bf a}\&{\bf b}, where the membrane is well resolved both for the quantum enhanced and shot-noise limited equivalent images. Fig.~\ref{image_comp}{\bf e} is the same plot in a region of the membrane where only the quantum enhanced image resolves the membrane with a signal-to-noise ratio higher than 1.

One way to illustrate the advantage provided by quantum correlations is to analyse the images for features that are not resolvable without these correlations. As can be seen by comparing the images in Figure~\ref{image_comp}{\bf a},{\bf b}~\&~{\bf c}, the semicircular  feature located vertically above main body of cell (consistent with a segment of the cell membrane) is significantly better resolved using quantum correlations than without them. To quantify this improvement we  estimate  the resolvable fraction of this feature for Figures~\ref{image_comp} {\bf a} and {\bf b}\footnote{We do not attempt to quantitatively compare the quantum enhanced image to Figure~\ref{image_comp}{\bf c} both because of the degradation in resolution in that image; and because switching the squeezed light source non changes the spatiotemporal profile of the Stokes beam making a quantitative comparison invalid (see Section~\ref{charact}).}. 
This is done by taking radial cross-sections of the images through the feature (referred to henceforth as the membrane), and interpolating the Raman signal along these cross-sections. The cross-sections show a small bump at the cell membrane (see Fig.~\ref{image_comp}{\bf d}~\&{\bf e}), which we define as resolvable if its maximum is more than a factor of two higher than the noise floor (equalling a signal-to-noise of 1). This yields an angular range for which the membrane signal is resolvable for each of the quantum-enhanced and shot-noise limited images. From this analysis, we found that the quantum-enhanced image allows a 40\% greater length of membrane to be resolved than the shot-noise limited image.   Assuming the membrane to be roughly circular, this corresponds to 23\% of the total membrane, compared to 16\% for the shot-noise limited image (shown angularly by the double-headed angles in Figs.~\ref{image_comp}{\bf a}~\&~{\bf b}).

\end{document}